\documentclass[letterpaper]{article} 
\usepackage{aaai25}  
\usepackage{times}  
\usepackage{helvet}  
\usepackage{courier}  
\usepackage[hyphens]{url}  
\usepackage{graphicx} 
\usepackage{natbib}  
\usepackage{caption} 
\usepackage{amsmath}
\usepackage{booktabs}
\usepackage{multirow}
\usepackage{xcolor}
\usepackage{placeins}
\usepackage{verbatim} 
\usepackage{xcolor}
\newcommand{\answerYes}[1]{\textcolor{blue}{#1}}

\newcommand{\answerNA}[1]{\textcolor{gray}{#1}}

\title{ChatGPT vs Teachers vs Students: Large-Scale Analysis of Generative AI Discourse in Education Communities on Reddit}

\author{
    Pelin Yüce\textsuperscript{\rm 1},
    Xiangruo Dai\textsuperscript{\rm 2},
    Rebecca Owens\textsuperscript{\rm 3},
    Tu\u{g}rulcan Elmas\textsuperscript{\rm 2}
}
\affiliations{
    \textsuperscript{\rm 1}Middle East Technical University\\
    \textsuperscript{\rm 2}The University of Edinburgh\\
    \textsuperscript{\rm 3}Durham University
}

\nocopyright
\begin{document}
\maketitle

\begin{abstract}
Generative Artificial Intelligence (GenAI) has prompted significant discussion in education, yet large-scale
empirical evidence on how students and teachers perceive and navigate this shift remains
limited. We analyse 270k AI-related Reddit posts and comments from 26 education-related
subreddits spanning higher education, K-12 teaching, and professional training between November 2022 and April 2026.
Topic modelling reveals seventeen themes covering \emph{academic integrity},
\emph{teaching \& pedagogy}, \emph{career anxiety}, \emph{policy}, and \emph{niche professional contexts}. Discourse evolves from an early detection-and-evasion arms race into a
sustained enforcement regime that constructive integration only begins to
challenge in mid-2024. Stakeholder communities differ sharply: K-12 teachers
foreground cognitive dependency, academics focus on AI detection and deliberation, and professional-programme students concentrate on career anxiety. Sentiment correlates strongly negatively with
engagement, showing adversarial enforcement themes mobilise communities far more than constructive integration discourse. Examining where faculty and students
meet, we find 17\% of threads are cross-role, and one third of such contact occurs in the adversarial themes \emph{AI Detection} and \emph{Misconduct Enforcement}.
Students initiate 68\% of mixed threads, but faculty produce most cross-role replies. Mixed threads contain 2-3$\times$ more records and last 2-4$\times$ longer than same-role threads, making adversarial integrity disputes the center of sustained faculty-student contact. We discuss implications for governance, pedagogical design, and cross-role contact design. The code and data is available at \url{https://github.com/tugrulz/genai-edu}
\end{abstract}

\section{Introduction}

Generative Artificial Intelligence (GenAI) tools have been transforming education since November 2022.
Students and staff alike can use AI tools for diverse tasks including language learning, writing and editing, and research and inquiries
\cite{fuchs2023exploring}. The accelerated adoption poses risks to academic integrity and
student learning: GenAI outputs are difficult to distinguish from student
work, complicating assessment and raising questions about the value of
education \cite{stohr2024perceptions, farazouli2024hello}.

This disruption has prompted policy recalibration, with universities revising
academic integrity frameworks, yet a striking gap
persists between governance structures regulating GenAI and the lived realities
of students and instructors. Prior empirical Reddit work \cite{wu2024reacting} addresses only the launch period and early reactions in HE-only subreddits, predating the normalisation of AI adoption and its spread to K--12 and professional training. A second gap concerns \emph{where stakeholders actually meet}: faculty and students inhabit largely separate subreddits and frame GenAI through different experiential lenses, making the rare cross-role threads a unique vantage on this contact. We address these gaps through a large-scale computational analysis of
education Reddit discourse. Specifically, we ask:

\begin{description}
    \item[RQ1:] What themes characterise online discussion of GenAI in education and how have these evolved?
    \item[RQ2:] How do these themes distribute across stakeholder communities, and how are theme prevalence, sentiment, and community engagement related?
    \item[RQ3:] Where and how do faculty and students meet in the same threads, which themes and communities host cross-role co-discussion, and how does that contact unfold?
\end{description}

Our contributions are threefold.
First, we present one of the largest Reddit-based analyses of GenAI in
education to date: 270k records from 26 subreddits spanning higher
education, K-12 teaching, and professional training over 3.5 years by a 17-theme taxonomy identified using LDA.
Second, we profile subreddits by theme, sentiment, and engagement, exposing how negativity and engagement co-vary across themes and subreddits.
Third, using LLM-based role classification of 100k users, we provide the first systematic analysis of where and how faculty and students meet in the same threads, establishing integrity disputes as the principal site of sustained cross-role contact about GenAI on Reddit.

\section{Related Work}


\noindent\textbf{GenAI Adoption in Education:} The ChatGPT launch marked a watershed in educational AI research. Research conducted shortly after ChatGPT's release identified benefits, including innovative pedagogy and learning design, as well as challenges, such as the novelty effect and ethical concerns~\cite{ren2025examining}. Adoption is now near-universal, with reported use by 92\% of UK students, up from 66\% in 2024~\cite{freeman2025student}, although engagement varies by discipline and gender~\cite{stohr2024perceptions}. 

Beyond these adoption-level findings, studies of AI use in practice reveal important student-faculty divides in attitudes and patterns of use. Students hold complicated attitudes and report mixed experience levels, nervousness over excitement, and pragmatic views of career and societal implications \cite{stone2025generative}. 

These concerns are reinforced by emerging evidence suggesting that AI reliance may reduce neural activity and impair learning outcomes~\cite{kosmyna2025your}, while LLM hallucinations pose risks because outputs may be unreliable or misleading~\cite{elsayed2024impact}. Trust and distrust operate as distinct rather than opposing dimensions, meaning that greater AI familiarity can support moderate trust in GenAI's capabilities while also sustaining high distrust of its risk~\cite{lyu2025understanding}. Both groups recognise AI's potential to enhance learning, with faculty reporting challenges with accuracy and integration, whilst students struggle with ethical application and reliability \cite{schmidt2025integrating}.


\noindent\textbf{Social Media Studies of GenAI in Education:} As advanced GenAI tools
produce text difficult to distinguish from human writing and changes assessment paradigms
\cite{stohr2024perceptions, farazouli2024hello}, academic integrity disputes have become salient in online education communities, motivating a growing body of
computational discourse analysis. Social media websites such as Reddit and Twitter are primary venues to capture such discourse as they can cover reactions to events~\cite{chausson2026beyond}, grievances~\cite{wang2026grievance}, cross-group contact~\cite{ccetinkaya2025cross} and communication patterns~\cite{bidewell2026gendered}. An early study by \citeauthor{wu2024reacting} (\citeyear{wu2024reacting}) found that 47.7\% of Reddit threads on academic AI included integrity discussions, with faculty focusing on AI usage and distrust in detection software, while students focused on false accusations. Subsequent computational studies corroborate and extend these findings.
Analysing 1,199 Reddit posts and 13,959 comments with sentiment analysis,
author classification, and LLM-assisted topic modelling, \citeauthor{devito2025unpacking}
(\citeyear{devito2025unpacking}) identify 12 discourse topics and confirm the
student--faculty asymmetry at a smaller scale. Both groups note productivity and learning benefits of Gen AI, but students emphasise false positives and detection issues, whereas educators focus on integrity, job security, and institutional policies.
\citeauthor{gaba2026groupdifferentiateddiscoursegenerativeai} (\citeyear{gaba2026groupdifferentiateddiscoursegenerativeai}) study 3,789 Reddit posts
across five subreddits and find teachers articulate explicit
pedagogical trade-offs (AI as simultaneously beneficial and harmful),
while students discuss AI tactically in terms of grades and enforcement.
Cross-platform evidence corroborates these themes: analysing Twitter data from April 2023 with topic modelling, \citeauthor{li2024chatgpt} (\citeyear{li2024chatgpt})
identify themes of academic integrity, learning outcomes,
tech limits, policy concerns, and workforce implications, which map closely onto our Reddit-derived 17-theme taxonomy.

We extend this body of work along four axes: corpus size and order
of magnitude beyond comparators (\citeauthor{devito2025unpacking},
\citeauthor{gaba2026groupdifferentiateddiscoursegenerativeai}); a longitudinal window covering three distinct
adoption phases rather than the launch wave alone; stakeholder coverage
that spans higher education, K--12, and professional training across four
Anglosphere countries rather than HE-only or single-country samples; and
a coherence-validated topic taxonomy rather than a heuristic or LLM-assisted
topic sets. Our findings depart from earlier reports of high positivity, which largely reflect the initial ChatGPT adoption wave rather than more settled conditions of GenAI use.\citeauthor{koonchanok2024public} (\citeyear{koonchanok2024public}), for example, analyse Twitter data from December 2022 to June 2023 finding broadly neutral-to-positive sentiment, alongside declining negative sentiment, which aligns with our Phase~1 Reddit findings, suggesting an early post-release “shock” period rather than a stable trajectory of public attitudes.

\section{Data \& Methods}


Using the API provided by Arctic Shift Project (\citeyear{arcticshift2022}), we
collected all posts and comments from 26 active education subreddits with more than 10,000 records, spanning the Anglosphere, covering November 2022 to April 2026 across six categories: faculty/HE, K-12 teachers, undergraduates, graduate students, professional programmes, and student support/edtech, detailed
in Table~\ref{tab:data}. Our collection has \textbf{22,352,353} total records
(1{,}744{,}457 posts and 20{,}607{,}896 comments).

\noindent\textbf{AI-Related Filtering:} Records enter the corpus if they match \emph{either} (1)~the bare-AI pattern
\texttt{\textbackslash b(AI|ai)\textbackslash b} or (2)~any named GenAI tool
pattern (Table~\ref{tab:patcounts}). The Bare-AI pattern captures hyphenated compounds
(\textit{AI-generated}, \textit{gen-AI}) as hyphens are non-word characters. Tools found by a preliminary analysis of Bare-AI records are matched case-insensitively: ChatGPT/GPT variants,
OpenAI, Claude, Gemini, Copilot, Perplexity, LLM, GenAI, Grammarly,
Quillbot, Turnitin, GPTZero/ZeroGPT/WinstonAI, and NotebookLM. The union of two patterns yields \textbf{270,929} records:
176,389 (65.1\%) bare-AI only;
62,468 (23.1\%) named-tool only and
32,072 (11.8\%) both.
Per-keyword counts and filter validation are documented in
Appendix~\ref{app:patcounts}.




\subsection{Topic Modelling}

\subsubsection{Preprocessing:} We cleaned each text by stripping URLs, Reddit markdown
syntax (asterisks, backticks, blockquote prefixes, and subreddit/user mentions), and
comments authored by AutoModerator or VettedBot (1,379 comments).
We removed records shorter than 15 tokens after cleaning
(Appendix~\ref{app:tokenhist}). For LDA, we further lemmatised texts with NLTK WordNet
and added bigrams (min\_count=10, threshold=15) via gensim Phrases, yielding a
38k-term vocabulary that we filtered to a 10k-term training vocabulary
($\text{min\_df}=5$, $\text{max\_df}=0.90$).

\begin{figure}[t]
\centering
\includegraphics[width=\columnwidth]{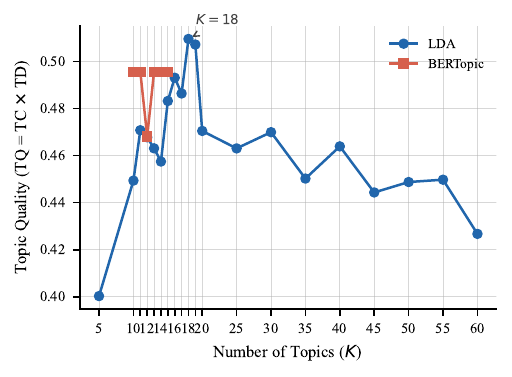}
\caption{Topic Quality (TQ $=$ TC$\,{\times}\,$TD) for LDA and BERTopic across $K$.
LDA peaks at $K=18$ (TQ$\,{=}\,0.509$)}
\label{fig:lda_k}
\end{figure}

\noindent\textbf{LDA:} 
We used Gensim's variational Bayes LDA with a symmetric Dirichlet prior
$\alpha = 1/K$ and $\eta = 1/K$, trained for 10 passes over the full corpus
with a fixed random seed (42) for reproducibility.
We selected $K$ by three complementary metrics:
topic coherence TC ($C_v$; \citealt{roder2015exploring}),
topic diversity TD (mean pairwise Jaccard of top-25 words),
and topic quality TQ $= \text{TC} \times \text{TD}$ \cite{dieng2020topic}.
A coarse sweep over $K \in \{5,10,\ldots,60\}$ identified a local peak near $K=15$ (TC$=0.507$);
a fine sweep over $K \in \{11,\ldots,19\}$ selected $K=18$
(TC$=0.528$, TQ$=0.509$), confirmed by a six-seed stability analysis (Appendix~\ref{app:opus_a1}).
Figure~\ref{fig:lda_k} reports all metrics.

\noindent\textbf{BERTopic:} As a complementary neural approach we tested BERTopic~\cite{grootendorst2022bertopic}
with \texttt{all-MiniLM-L6-v2} embeddings, UMAP, and HDBSCAN, with topic
representations refined via KeyBERT re-ranking, MMR, and POS filtering
(full configuration in Appendix~\ref{app:bertopic}). HDBSCAN converged to only two stable density states regardless of the requested $K$: five effective topics
($K \in \{10,11,13,14,15\}$; TC$=0.544$) or eleven ($K=12$; TC$=0.519$).
BERTopic, therefore, serves as convergent validation; topic-level correspondence
with LDA $K=18$ is in Appendix~\ref{app:bertopic}.

\noindent\textbf{Topic Assignment:} We assigned each document to its dominant LDA topic and manually labelled the topics by inspecting the top-25 words and representative posts. We then grouped the 18 topics into 5 thematic clusters based on this inspection to facilitate downstream analysis.

\noindent\textbf{Human Validation:} We randomly sampled 583 posts for coding by employing a minimum of 3 posts per half year (8 half years in total) per topic constraint. Two annotators independently coded the data by reviewing and, where necessary, correcting the existing topic assignments. Inter-annotator agreement was fair at the topic level
(Cohen's $\kappa{=}0.38$, raw agreement 43.4\%) and moderate at the
five-cluster level ($\kappa{=}0.53$, 62.3\%); LDA-vs-human agreement
was comparable (topic $\kappa{=}0.38$--$0.51$; cluster
$\kappa{=}0.45$--$0.56$). The largest sources of topic-level
disagreement are within-cluster boundary cases
(e.g., \emph{Personal Misconduct Narratives} vs.\
\emph{Misconduct Enforcement} vs.\ \emph{AI Detection \& False
Accusations}, which together comprise the \emph{Academic Integrity}
cluster). We therefore anchor the paper's interpretation
at the five-cluster level. Full inter-annotator and LDA-vs-human
contingency analysis is in Appendix~\ref{app:opus_b8}.

\noindent\textbf{Relevance:}
\label{sec:relevance}
The keyword filter admits a small fraction of records that mention
``AI'' or ``LLM'' without engaging substantively with GenAI. We employed Llama 3.3 70B to automatically classify
records with a \emph{weak} AI signal (bare-AI-only matches, plus
LLM-only named-tool matches that might be \emph{Master of Laws}).
Relevance can be verified either \emph{before} fitting
LDA (filter the corpus, then model) or \emph{after} (model the full
corpus, then verify each topic's representative records). We found
that $K=18$ provides the highest TQ in both cases. Verifying after
topic modelling lets a single annotation pass both vet candidate
themes and flag irrelevant records and is cheaper than running two
separate annotation rounds. Thus, we adopt the post-hoc order.
Llama flagged 17,707 records ($\sim$6.5\% of the corpus) as not relevant; we remove these and continue the study with the remaining 253,222 records (system prompts and full filtering details in Appendix~\ref{app:opus_b8}).
To validate this classifier, we compared Llama's relevance labels
against the two-annotator topic-coding sample (n${=}583$), in which
both annotators marked records
as \emph{irrelevant} if they do not substantively engage with GenAI in education
(Annotator B: 26\%; Annotator A: 10\%; both-agreed: 9\%; inter-annotator
$\kappa{=}0.42$). Llama agreed with
the more conservative annotator B at $\kappa{=}0.50$ (84.6\% raw
agreement) and with the both-agreed consensus at $\kappa{=}0.49$, above the human inter-annotator ceiling. 

Across the 18 LDA topics, 12 retain over 92\% relevance after the
Llama filter. Of the remaining 6,
four are career-related (T07, T09, T12, T14) with 12--19\% irrelevant
posts driven by AI-adjacent job and degree discussions outside our
scope; one (T04) reaches above 50\% irrelevance because ``LLM'' in
r/LawSchool also denotes \emph{Master of Laws}; and T11 (1,754
records, 0.7\%) is excluded entirely as an incoherent r/HomeworkHelp
spillover cluster (Prolog puzzles, physics, thermodynamics) flagged
by both human and AI validation. The LLM system prompts,
per-topic and per-subreddit relevance shares, audit results, and the
post-filter $K$-sweep are in Appendix~\ref{app:opus_b8}
(Table~\ref{tab:opus_b8_ksweep}).

\noindent\textbf{Sentiment Analysis:} We used \texttt{twitter-\allowbreak roberta-\allowbreak base-\allowbreak sentiment-\allowbreak latest}~\cite{barbieri2020tweeteval}, a scalable 125M-parameter model trained on social media data, to compute a per-record sentiment score $s = P_{+} - P_{-} \in [-1, +1]$.




\subsection{Role Classification}
\label{subsec:role-detection}

We inferred the role (\textsc{Faculty} or \textsc{Student}) for each
author using Llama~3.3~70B~Instruct \cite{grattafiori2024llama} covering all 109,581 authors in the corpus. We showed the LLM up to five post/comment bodies per author
(subreddit names, usernames, and metadata hidden to avoid bias) and requested one of four
labels: \textsc{Faculty}, \textsc{Student}, \textsc{Dual} (e.g.\ a graduate
teaching assistant), or \textsc{Unclear} (insufficient evidence). See Appendix~\ref{app:roledetection} for the LLM prompt and experimental details.

Of 109,581 authors, 82,373 received a binary label 75\% of authors, covering 81\% of the corpus), 35,466 \textsc{Faculty} and 46,907 \textsc{Student}.
A further 1,910 authors received \textsc{Dual} and 25,298 \textsc{Unclear} and
were excluded from all role-stratified analyses. The full methodology is in Appendix~\ref{app:roledetection}.

\noindent\textbf{Human Validation of Role Labels:}
\label{subsec:annotation-validation} To assess the validity of the automated role-classification pipeline,
two expert annotators independently labelled a stratified random sample
of 200 users drawn from the comprehensive role assignment table.
The sample was balanced across two strata: 100 \textit{subreddit-clear} users who only post to subreddits that are self-labelled with a role (e.g., users who post only to r/Professors) and 100 \textit{challenging} users whose role is unclear from their subreddit activity. Annotators were shown the same post/comments given to the classifiers, subreddit names, and assigned one of four labels.

The inter-annotator agreement is substantial overall (Table~\ref{tab:llama_agreement}), reaching $\kappa = 0.603$ (73.0\%) and rising to near-perfect levels under binary restriction ($\kappa = 0.887$, 94.3\%). Agreement between Llama and the human annotators is moderate but approaches human levels on clearer instances (e.g., $\kappa = 0.447$ for Annotator A on the subreddit-clear set) and improves when restricted to binary labels, mirroring the same pattern observed in human agreement.

\begin{table}[t]
\centering
\small
\setlength{\tabcolsep}{4pt}
\begin{tabular}{lcccccc}
\toprule
Set & \multicolumn{2}{c}{A vs Llama} & \multicolumn{2}{c}{B vs Llama} & \multicolumn{2}{c}{A vs B} \\
 & $\kappa$ & (\%) & $\kappa$ & (\%) & $\kappa$ & (\%) \\
\midrule
Overall  & 0.383 & 57.0 & 0.433 & 63.0 & 0.603 & 73.0 \\
Binary-only  & 0.364 & 56.1 & 0.463 & 66.7 & 0.887 & 94.3 \\
Subreddit-clear & 0.447 & 62.0 & 0.433 & 62.0 & 0.658 & 77.0 \\
Challenging & 0.333 & 52.0 & 0.433 & 64.0 & 0.545 & 69.0 \\
\bottomrule
\end{tabular}
\caption{Agreement between Llama and annotators. Binary-only excludes \textsc{Unclear} and \textsc{Dual}.}
\label{tab:llama_agreement}
\end{table}

\subsection{Compute}
LDA ran on CPU
($\sim$2.3 h); BERTopic and RoBERTa sentiment scoring ran on a local NVIDIA RTX 4060 ($\sim$1.5 h);
Llama 3.3 70B ran via an
institutional inference endpoint ($\sim$39 h batched API calls). Relevant licences in Appendix~\ref{app:licences}.
\section{Results}
\label{sec:results}

\subsection{RQ1: Discourse Themes and Their Evolution}

We manually inspect the top-25 words, top 5 posts, and posts sampled in
human validation to name and describe each topic. Table~\ref{tab:lda_topics} lists all seventeen themes with corpus shares and top keywords.
The Fac\% column reports the share of records (posts and comments, not unique authors) in each topic
attributed to faculty-identified users, showing the degree to which themes are faculty- or
student-driven. Though some themes appear similar, sample inspection confirms all
seventeen topics are distinct. We now describe the 17 themes grouped into 5 overarching thematic clusters, highlighting the boundary distinctions between adjacent themes.

\begin{table*}[!t]
\centering
\scriptsize
\setlength{\tabcolsep}{3pt}
\renewcommand{\arraystretch}{0.9}
\begin{tabular}{p{1.2cm}cp{3.9cm}rrp{7.4cm}rr}
\toprule
\textbf{Cluster} & \textbf{ID} & \textbf{Label} & \textbf{N} & \textbf{\%} & \textbf{Description} & \textbf{Fac\%} & \shortstack[r]{\textbf{Cmts}\textbf{/post}} \\
\midrule
\multirow{4}{*}{\shortstack[l]{Academic\\Integrity\\(37.1\%)}}
 & T05 & Misconduct Enforcement & 30,353 & 12.1 & Faculty/TA detecting, confronting, and grading AI-generated work; appeals & 74.8 & \textbf{35.5} \\
 & T02 & AI Detection \& False Accusations & 27,275 & 10.8 & Turnitin/detector use; false-positive accusations; student appeals & 54.1 & 17.1 \\
 & T00 & Personal Misconduct Narratives & 19,556 & 7.8 & First-person stories of AI accusation, punishment, and institutional appeal & 56.3 & \textbf{24.9} \\
 & T17 & AI Writing Quality \& Evasion & 16,051 & 6.4 & Recognising AI prose; evasion strategies; echowriting; detection avoidance & 66.8 & \textbf{21.9} \\
\midrule
\multirow{4}{*}{\shortstack[l]{Teaching \&\\Pedagogy\\(28.0\%)}}
 & T16 & Frontline AI Reactions \& Opinions & 21,841 & 8.7 & Short emotional responses to AI in daily classroom life; colloquial register & 60.4 & \textbf{37.5} \\
 & T13 & AI-Assisted Workflow \& Help-Seeking & 21,159 & 8.4 & Task-completion AI use: IEP goals, lesson plans, emails, study aids & 51.7 & 10.6 \\
 & T08 & Assessment Redesign \& Teaching & 13,907 & 5.5 & Faculty redesigning exams and assignments in response to AI & 73.1 & \textbf{22.7} \\
 & T01 & Learning Quality \& Cognitive Dep. & 13,421 & 5.3 & AI eroding critical thinking, problem-solving, and core learning outcomes & 82.3 & 15.0 \\
\midrule
\multirow{3}{*}{\shortstack[l]{Career \&\\Future\\Anxiety\\(15.7\%)}}
 & T12 & Career \& Personal Economic Anxiety & 25,767 & 10.2 & Personal career decisions under AI threat; degree value; AI-proof fields & 52.4 & \textbf{21.4} \\
 & T09 & Degrees, Programs \& Graduate Study & 11,417 & 4.5 & Degree choice, programme value, and graduate applications in the AI era & 19.6 & 4.4 \\
 & T07 & AI Job Displacement & 2,304 & 0.9 & Macro-societal fear of AI replacing entire professions; abstract/ideological & 64.2 & 8.0 \\[4pt]
\midrule
\multirow{3}{*}{\shortstack[l]{Policy \&\\Deliberation\\(13.2\%)}}
 & T10 & Deliberative AI Discourse & 26,906 & 10.7 & Analytical debate on AI capabilities, ethics, and policy; technical register & 71.0 & 17.1 \\
 & T06 & Research, Publishing \& GenAI Ethics & 5,347 & 2.1 & AI in literature reviews, journal submissions, citation fabrication, authorship & 62.0 & 12.5 \\
 & T04 & Institutional Policy \& Legal & 953 & 0.4 & Institutional AI policies, legal frameworks, professional liability & 56.4 & 10.1 \\
\midrule
\multirow{3}{*}{\shortstack[l]{Niche \&\\Professional\\(6.0\%)}}
 & T03 & AI Tool Selection \& Features & 10,703 & 4.3 & Comparing specific AI tools by features, cost, and suitability & 58.6 & 4.7 \\
 & T14 & Job Applications \& Professional & 3,610 & 1.4 & AI-drafted cover letters, recommendation letters, and student emails & 51.9 & 5.7 \\
 & T15 & Healthcare \& Medical Education & 895 & 0.4 & AI in clinical training; replacement fears; ChatGPT for USMLE study & 55.2 & 9.0 \\
\bottomrule
\end{tabular}
\caption{All themes grouped into five thematic clusters (sorted by cluster
share, then by topic size within each cluster).
Fac\% = faculty/(faculty+student) records per topic.
Cmts/post = mean comments per post; \textbf{bold} = $\geq$20 (high-engagement topic).}
\label{tab:lda_topics}
\end{table*} 

\noindent\textbf{Academic Integrity (37.1\%).} The largest cluster, accounting for more than a third of all AI-related records, spans the full detection--enforcement--evasion cycle.
The three event-centric topics of \emph{Misconduct Enforcement}, \emph{AI Detection \& False Accusations}, and \emph{Personal Misconduct Narratives} form a ``cheating triangle'' in our two-annotator validation: 64 of 226 substantive disagreements (28\%) fall on these three pairs, reflecting that detection, faculty enforcement, and student narratives are three vantages on the same incident rather than independent themes. \emph{AI Writing \& Evasion} sits more loosely: it contains discussion on both detection and ethics in the form of ``legitimate" use of AI-assisted writing.
\textit{Misconduct Enforcement} (12.1\%) is the single largest theme: teaching staff share burdens of enforcement dilemmas, evidentiary standards, and grade appeals without institutional frameworks or reliable tools. The most upvoted post in the corpus (22,888 upvotes) is from a teacher who recommends using hidden white-text prompts in assignment documents to identify AI-assisted student submissions.
\textit{AI Detection \& False Accusations} (10.8\%) splits between faculty using Turnitin to flag cheating and students contesting false positives, revealing a legitimacy crisis. Our sample analysis shows that detectors are reported as linguistically biased, unreliable, and psychologically damaging. The most upvoted detection posts are all from students; one of the top posts (3,292 upvotes) concerns a student whose original assignments were falsely flagged as AI-generated by detection software. 
\textit{Personal Misconduct Narratives} (7.8\%) contains confessions (or denials) of misconduct and accounts of students' accusation, denial and anxiety, institutional appeal and sanctions. Stories of injustice or moral failure generate some of the highest engagement. There is some overlap with \emph{AI Detection \& False Accusations}, with some students pleading innocence and discussing experiences with academic integrity boards after the false accusations. 
\textit{AI Writing Quality \& Evasion} (6.4\%) contains faculty recognising signs of AI usage, including prose and vocabulary going against students evading detection through evolved prompting and ``echowriting'' (paraphrasing ChatGPT output); hallucinated citations and cliche phrases are a recurring sub-thread.

\noindent\textbf{Teaching \& Pedagogy (28.0\%).}
This cluster covers the day-to-day classroom encounter with AI across emotional, practical, and pedagogical dimensions.
\textit{Frontline AI Reactions \& Opinions} (8.7\%) comprises short, colloquial posts in r/Teachers and r/CollegeRant (keywords including \emph{kids, lazy, slop, stupid, cheat}); main themes are defeatism, open discussion \textit{``Are modern students really that far behind, or is it overexaggerated?''} and acceptance. Averaging 54 words, this theme has the highest comments-per-post (37.5), and includes highly engaged discussions with emotional expressions such as ``tired,'' ``frustrated,'' ``horror,'' ``sad,'' ``hate,'' and ``hope''.
\textit{AI-Assisted Workflow \& Help-Seeking} (8.4\%) is more everyday exposition and solution-oriented. In K-12 contexts, discussions revolve around administrative tasks such as lesson planning, parent emails, and disability-related adjustments, while in higher education, they focus on learning support and practical research uses. There is accordingly lower emotional engagement despite the high number of posts, with few posts receiving over 50 votes.
\textit{Assessment Redesign \& Teaching Innovation} (5.5\%) centers on both questions on how teaching/assessment practices and pedagogy are being reconfigured to counter the AI era, including faculty recommending return to closed book exam, redesigning assessments, and creatively AI-proofing methods; a top post on open-book exams producing unexpectedly high failure rates exemplifies the instructor's frustration with AI's impact on traditional assessment methods. 

\textit{Learning Quality \& Cognitive Dependency} (5.3\%) addresses the deeper question of the effects of AI use on students' abilities, including whether it erodes critical thinking, problem-solving, and coding skills. Posts from educators describing students' loss of programming competence are prominent, as are discussions of over-reliance on core learning tasks, with concerns concentrated in faculty-dominated r/Teachers and r/Professors. A top post from the latter bemoans \emph{``we need to stop pretending the house isn't on fire while we're repainting the walls''} and the former \emph{``AI is the gateway drug that will end critical thinking''}, reflecting widespread discontent with students' learning abilities.

\noindent\textbf{Career \& Future Anxiety (15.7\%).}
Three themes share underlying anxiety about AI's effect on human work, operating at different levels of society.
\textit{Career \& Personal Economic Anxiety} (10.2\%) shows individual rather than societal fears, questioning which degrees are AI-proof.
This discussion is prominent in r/LawSchool, r/medicalschool, and r/nursing, where professional identity is tightly bound to credentials. However, not all posts are pessimistic; a top post in r/nursing doubted their department would buy AI tools when it couldn't even \textit{``replace a 9.5yr old PC.''} {Comments under such posts largely revolve around concerns of partial displacement instead of full replacement.
Themes include workload intensification and labour reassignment following integration of AI tools into educational/clinical workflows. Users frequently express concerns that AI would reduce staffing levels, reduce autonomy, and leave health professionals with more physically and emotionally demanding tasks. One commenter noted that \emph{“[AI] will handle the easy stuff, [humans] delicate, more stressful situations.”}
\textit{Degrees, Programs \& Graduate Study} (4.5\%) include analysis of career choices and outcomes \textit{``[PhD applicants, know that] Stanford's faculty in CS are leaving,''} \textit{``I've \pounds 90k in student debt [with no job due to AI]''} and generate the lowest comments-per-post of any theme (4.4), suggesting less engagement due to lowered personal investment.

\textit{AI Job Displacement} (0.9\%), though related to \emph{Career \& Personal Economic Anxiety}, considers societal-level career threats and speculation of AI replacing entire professions. Despite its prevalence in mainstream media, it remains a niche topic in the corpus. Discourse ranges from downplaying AI (\textit{``AI Bubble is going to burst soon''}) to structural critique \textit{`` [despite AI]... schools will still exist because you need childcare''}. Despite being negative in sentiment, it generates little engagement, indicating that the personal stakes of \emph{Career \& Personal Economic Anxiety} are more appealing.

\noindent\textbf{Policy \& Deliberation (13.2\%).}
This cluster contains the most institutionally oriented discourse.
\textit{Deliberative AI Discourse} (10.7\%) uses technical vocabulary (\emph{LLM, inference, training, FERPA, Bayesian}), produces long comments (mean 95 words, nearly double the corpus mean), and concentrates in r/PhD, r/AskAcademia, r/LawSchool, and r/Academia. Recurring sub-threads debate the nature of ChatGPT, data-governance obligations, and the ethics and reliability of AI in research. Although \emph{Deliberative AI Discourse} and \emph{Frontline AI Reactions \& Opinions} share general-AI-discourse vocabulary, the former is analytical while the latter is more emotional. 
\textit{Research, Publishing \& GenAI Ethics} (2.1\%) addresses the proper usage and ethics of LLMs in literature reviews, AI-generated journal submissions, and authorship declaration; smaller in volume but disproportionately concentrated in r/PhD and r/Academia. Topics included discussion of appropriate disclosure for AI usage in research, complaints about widespread use of AI amongst colleagues, and an academic \emph{``tired of dealing with slop as a reviewer."}

\textit{Institutional Policy \& Legal Frameworks} covers school and module AI policies, and legal liability; r/LawSchool is the largest single contributor (34\% of theme records, followed by r/Teachers at 14\% and r/Professors at 11\%), focusing on professional-consequence discourse (attorney sanctions for using ChatGPT in briefings, hallucinated citations in filings), and a post disclosing a mandate that \emph{``instructors may not prohibit AI use"} generated significant backlash, suggesting an emerging fault line between institutional policymakers and faculty.

\noindent\textbf{Niche \& Professional (6.0\%).} \textit{AI Tool Selection \& Features} (4.3\%) compares tools by capability and cost (Claude vs.\ DeepSeek vs.\ ChatGPT; coding assistants), concentrated in r/PhD, r/edtech, and r/GradSchool. The posts ask \emph{which tool?} rather than \emph{how do I use AI for this task?} common in \emph{AI-Assisted Workflow \& Help-Seeking}. Many posts are written by marketers promoting their own software; phrases such as \emph{I created} and \emph{I built an AI tool} are common.

\textit{Job Applications \& Professional Communication} (1.4\%) covers AI-drafted CVs, recommendation letters, and student emails, with top posts including faculty frustration withformulaic AI emails from students asking to enrol in full courses and requesting extensions.

\textit{Healthcare \& Medical Education} (0.4\%) is heavily concentrated in r/nursing, r/medicalschool, and r/premed; concerns include AI hallucinating clinical information, replacing nurses and doctors, and AI-created flashcards. The emotional discourse shows that, AI-induced or not, clinical errors and job displacement are high-stakes.
Although we group \emph{Healthcare \& Medical Education} under \emph{Niche \& Professional} for compositional reasons (its discourse is bounded by a small set of clinical subreddits), the substantive content overlaps heavily with the \emph{Career \& Future Anxiety} cluster as indicated by human validation. 

\subsection{Discourse Theme Evolution}

\begin{figure*}[!t]
\centering
\includegraphics[width=\textwidth]{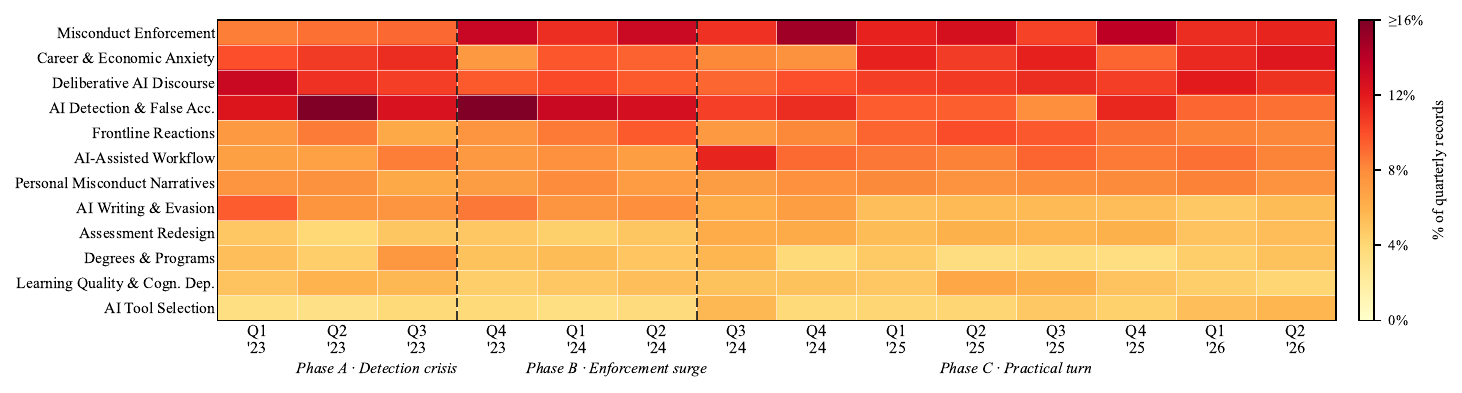}
\caption{Quarterly share (\%) of each discourse theme. Dashed white lines mark content-based phase boundaries. Warmer colours indicate higher monthly prevalence. Small themes are omitted due to data sparsity
}
\label{fig:topictime}
\end{figure*}

We analyse the temporal evolution of the themes. To facilitate the interpretation of quarterly changes in the volume of themes, we employ thematic change-point detection algorithms on the 17-dimensional monthly topic-proportion to identify 3 phases of the discourse.  We employ Binary Segmentation (BinSeg)
and Dynamic Programming (Dynp) changepoint algorithms \citep{truong2020selective}
to the 17-dimensional monthly topic-proportion vector, under both $\ell_2$
and RBF kernel costs.
 Three of the four methods--cost
combinations (BinSeg+$\ell_2$, BinSeg+RBF, Dynp+$\ell_2$) agree on the
same two-breakpoint solution (Full details in Appendix~\ref{app:opus_b2}). The phases are named by the dominant
compositional signal that drives the boundary or characterises the segment. Figure~\ref{fig:topictime} maps the quarterly share of each discourse theme
from Q1 2023 to Q1 2026.

\noindent\textbf{Phase~A: Detection crisis} (Nov 2022--Aug 2023).
\emph{AI Detection \& False Accusations} averages 14.1\% across Phase~A
(peaking at 21.6\% in November 2022 immediately after ChatGPT's launch)
and \emph{AI Writing Quality \& Evasion} peaks at 12.3\%.
The community is preoccupied with the detection/evasion arms race
immediately following ChatGPT's launch.

\noindent\textbf{Phase~B: Enforcement surge} (Sep 2023--Jun 2024).
Named for the single largest compositional shift across any topic at any
boundary: \emph{Misconduct Enforcement} surges from an 8.8\%
Phase~A mean to 12.5\% in Phase~B ($+$3.7~pp),
statistically driving the September~2023 breakpoint.
The community transitions from detecting AI to managing institutional
consequences.
\emph{Career Anxiety} peaks at 11.4\% in Q3~2023 ($+$1.2~pp above the Phase~A mean) before being displaced by enforcement discourse; \emph{AI Detection} amplifies to 16.3\% in Q4~2023, possibly driven by autumn midterms and finals.

\noindent\textbf{Phase~C: Practical turn} (Jul 2024--Apr 2026).
Three constructive themes gain share simultaneously relative to Phase~A:
\emph{AI-Assisted Workflow} ($+$2.1~pp), \emph{Assessment Redesign}
($+$1.4~pp), and \emph{Tool Selection} ($+$1.1~pp), while \emph{AI
Detection} falls by 4.4~pp.
The community moves from policing AI to integrating it.
\emph{AI-Assisted Workflow} reaches 11.5\% in Q3~2024: the practical turn opens with a constructive burst that moderates over subsequent quarters.
\emph{Misconduct Enforcement} spikes to 15.1\% in Q4~2024 ($+$2.7~pp), enforcement reasserting itself at the start of the 2024--25 academic year.

\subsection{RQ2:  Communities, Sentiment, and Engagement}


\begin{table*}[!t]
\centering
\scriptsize
\setlength{\tabcolsep}{2pt}
\renewcommand{\arraystretch}{0.9}
\begin{tabular}{p{1.2cm}p{2.2cm}rrrrrp{9.2cm}r}
\toprule
\textbf{Category} & \textbf{Subreddit} & \textbf{Posts} & \textbf{Cmts} & \shortstack[r]{\textbf{AI }\textbf{Posts}} & \shortstack[r]{\textbf{AI }\textbf{Cmts}} & \textbf{AI\%} & \textbf{Top-3 discourse themes} & \textbf{Fac\%} \\
\midrule
\multirow{6}{*}{\shortstack[l]{Faculty /\\Higher Ed.}}
 & r/Professors & 52K & 1.4M & 4,361 & 58,604 & 24.9 & Misconduct Enforcement (22\%), Deliberative AI Discourse (12\%), Assessment Red. (10\%) & 90.6 \\
 & r/AskAcademia & 47K & 476K & 1,604 & 8,869 & 4.1 & Deliberative AI Discourse (17\%), AI Detection (10\%), AI-Assisted Workflow (9\%) & 60.7 \\
 & r/Academia & 21K & 221K & 1,149 & 7,396 & 3.4 & AI Detection (17\%), Deliberative AI Discourse (17\%), AI-Assisted Workflow (8\%) & 63.1 \\
 & r/AskProfessors & 13K & 189K & 527 & 5,180 & 2.3 & Misconduct Enforcement (22\%), AI Detection (18\%), Deliberative AI Discourse (12\%) & 70.5 \\
 & r/education & 39K & 233K & 817 & 5,524 & 2.5 & Learning Quality (20\%), Frontline Reactions (12\%), Deliberative AI Discourse (10\%) & 68.5 \\
 & r/highereducation & 3.4K & 35K & 77 & 444 & 0.2 & Learning Quality (18\%), Deliberative AI Discourse (15\%), Career Anxiety (14\%) & 80.9 \\
\midrule
\multirow{3}{*}{\shortstack[l]{K-12 /\\Teachers}}
 & r/Teachers & 200K & 4.7M & 3,428 & 32,878 & 14.3 & Frontline Reactions (14\%), Misconduct Enforcement (13\%), Learning Quality (12\%) & 80.0 \\
 & r/teaching & 24K & 452K & 438 & 6,094 & 2.6 & Learning Quality (15\%), AI-Assisted Workflow (12\%), Misconduct Enforcement (11\%) & 85.2 \\
 & r/TeachingUK & 24K & 286K & 154 & 2,048 & 0.9 & AI-Assisted Workflow (20\%), Deliberative AI Discourse (12\%), Career Anxiety (10\%) & 88.2 \\
\midrule
\multirow{6}{*}{Undergrad}
 & r/UniUK & 110K & 1.2M & 2,410 & 18,454 & 8.2 & AI Detection (24\%), Career Anxiety (10\%), Deliberative AI Discourse (10\%) & 33.2 \\
 & r/College & 152K & 1.3M & 1,479 & 10,440 & 4.7 & AI Detection (22\%), Misconduct Enforcement (15\%), Career Anxiety (11\%) & 28.3 \\
 & r/CollegeRant & 30K & 295K & 709 & 8,979 & 3.8 & Misconduct Enforcement (23\%), AI Detection (19\%), Frontline Reactions (11\%) & 36.7 \\
 & r/University & 24K & 50K & 924 & 1,623 & 1.0 & AI Detection (28\%), AI-Assisted Workflow (12\%), Misconduct Enforcement (8\%) & 26.8 \\
 & r/UKUniversityStudents& 11K & 25K & 218 & 311 & 0.2 & Degrees \& Programs (22\%), AI Detection (18\%), Career Anxiety (13\%) & 22.2 \\
 & r/CanadaUniversities & 7.9K & 43K & 143 & 269 & 0.2 & Degrees \& Programs (38\%), Career Anxiety (21\%), AI-Assisted Workflow (7\%) & 20.0 \\
\midrule
\multirow{3}{*}{Graduate}
 & r/PhD & 59K & 781K & 2,247 & 14,127 & 6.5 & Deliberative AI Discourse (14\%), AI-Assisted Workflow (13\%), Degrees \& Programs (13\%) & 42.2 \\
 & r/GradSchool & 63K & 368K & 1,069 & 5,986 & 2.8 & AI Detection (18\%), Misconduct Enforcement (10\%), Deliberative AI Discourse (10\%) & 38.6 \\
 & r/gradadmissions & 158K & 76K & 3,869 & 174 & 1.6 & Degrees \& Programs (77\%), Career Anxiety (5\%), Job Applications (5\%) & 1.8 \\
\midrule
\multirow{6}{*}{Professional}
 & r/LawSchool & 86K & 1.0M & 1,046 & 8,344 & 3.7 & Career Anxiety (18\%), Deliberative AI Discourse (18\%), AI-Assisted Workflow (12\%) & 35.2 \\
 & r/medicalschool & 105K & 1.3M & 1,015 & 8,892 & 3.9 & Career Anxiety (31\%), Deliberative AI Discourse (16\%), Frontline Reactions (10\%) & 35.2 \\
 & r/premed & 162K & 1.3M & 711 & 4,024 & 1.9 & Career Anxiety (21\%), AI Writing \& Evasion (13\%), Personal Misconduct (11\%) & 17.0 \\
 & r/nursing & 188K & 3.9M & 635 & 6,709 & 2.9 & Career Anxiety (32\%), Personal Misconduct (13\%), Frontline Reactions (12\%) & 44.8 \\
 & r/StudentNurse & 44K & 363K & 156 & 1,685 & 0.7 & AI-Assisted Workflow (27\%), Assessment Redesign (23\%), Personal Misconduct (9\%) & 16.5 \\
 & r/DentalSchool & 18K & 143K & 72 & 328 & 0.2 & Career Anxiety (28\%), AI-Assisted Workflow (19\%), Tool Selection (13\%) & 27.0 \\
\midrule
\multirow{2}{*}{Other}
 & r/edtech & 5.8K & 24K & 1,203 & 3,504 & 1.9 & Learning Quality (23\%), AI-Assisted Workflow (20\%), Tool Selection (20\%) & 78.6 \\
 & r/HomeworkHelp & 97K & 398K & 563 & 1,278 & 0.7 & Personal Misconduct (17\%), Deliberative AI Discourse (16\%), AI-Assisted Workflow (14\%) & 26.5 \\
\bottomrule
\end{tabular}
\caption{Subreddit statistics under R3 thread-propagated relevance (253,222 records retained; Methods~\S\ref{sec:relevance}). AI\% = share of all R3-filtered AI records; Fac\% = faculty share of role-classified records.}
\label{tab:data}
\end{table*}

Table~\ref{tab:data} shows each subreddit's AI-related record counts,
top-3 themes, and faculty percentage. We next describe the subreddits with notable patterns in role and theme composition.

\noindent\textbf{r/Professors:} the most faculty-voiced community in the corpus (Fac\%~=~90.6), leading with \emph{Misconduct Enforcement} (22\%), \emph{Deliberative Discourse} (12\%), and \emph{Assessment Redesign} (10\%).
Many top posts address the burden of managing academic integrity cases, poor institutional support, and burnout, as well as cheating methods. These themes show a shift to systematic management of enforcement and assessment redesign. Students posting here often do so under \emph{AI Detection \& False Accusations} including appealing (and sometimes attacking) faculty, showing clear conflicts between student and academic interests in AI usage. In contrast to the /Teachers community, here AI is primarily encountered through submitted assessments and institutional policies, resulting in a focus on \emph{Deliberative Discourse} and \emph{Assessment Redesign}.



\noindent\textbf{r/Teachers} (Fac\%~=~80)\textbf{:} more distributed across \emph{Frontline Reactions} (14\%), \emph{Misconduct Enforcement} (13\%), and \emph{Learning Quality} (12\%). Almost all of the top posts in Learning Quality come from this community. Common themes include worries about student abilities and the potential decline due to technology, including AI, such as basic writing and reading skills. \emph{Misconduct Enforcement} rotates around sharing tips on detecting AI rather than purely imposing disciplinary sanctions on students for using it. This reflects K-12 educators' pedagogy and close observation of AI's impact on learning, making skill erosion and cognitive deficiencies a more immediate concern for them. 



\noindent\textbf{r/College, r/UniUK, and r/GradSchool:} all lead with \emph{AI Detection \& False Accusations} (22\%, 24\%, 18\%) mainly driven by students (Fac\% 28--39\%). Common topics include false flagging by Turnitin and credential values. Students care deeply about institutional policies and sanctions for improper AI usage. Simultaneously, the flood of discussion about AI detection leads to a widely shared \textit{metapost} urging others to stop discussing it. r/UKUniversityStudents is an exception as it is uniquely led by \emph{Degrees \& Programs} (22\%), with \emph{Career Anxiety} also prominent (13\%). Credential-related concerns thus account for 35\% of its AI discourse, possibly due to a higher applicant ratio, similar to /CanadaUniversities, in the prominence of degree-related posts, rather than to differences between British and other Anglosphere students.





\noindent\textbf{r/LawSchool, r/medicalschool, and r/nursing:} mainly student driven communities (Fac\%~35-45\%). In r/medicalschool and r/nursing \emph{Career Anxiety} dominates at 31\% and 32\% respectively; in r/LawSchool, \emph{Career Anxiety} (18\%) and \emph{Deliberative AI Discourse} (18\%) are essentially tied, with legal-AI discussion split between professional-anxiety threads and deliberation about AI capabilities and ethics. The high share of \emph{Career Anxiety} may signal perceived replaceability within credentialised professions. Rather than focusing on imminent replacement by technology, top posts often poke fun at AI's weaknesses, including its price and inability to perform higher-level tasks and sarcastic analyses of AI's failures are common.




\noindent\textbf{r/AskAcademia and r/Academia:} both lead with \emph{Deliberative Discourse} ($\sim$17\%; in r/Academia tied with \emph{AI Detection}). By name /AskAcademia is a student/non academic initiated community, and is reflected in the balanced faculty-student composition and genuine cross-role exchange. Top posts include AI's influence on degree values and careers, including one poster who asked if given AI's advances it was \emph{Time to leave academia - the fate of all applied fields}.







\noindent\textbf{r/edtech:} due to its topic, it proportionally contains the most AI-related discussion (16\% of all records). Topics include \emph{Learning Quality} (23\%), \emph{AI-Assisted Workflow} (20\%), and \emph{Tool Selection} (20\%), consistent with practitioners and developers promoting AI integration in teaching. Many posts are promotion/spam by ed-tech creators and marketers, making students and faculty an audience rather than contributors.



\begin{figure}[!t]
\centering
\includegraphics[width=\columnwidth]{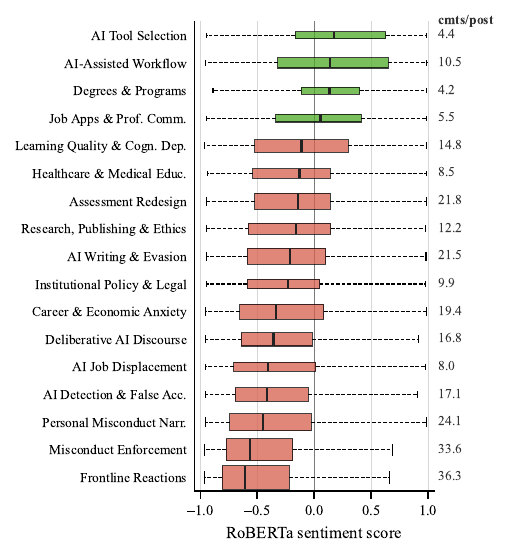}
\caption{Sentiment score per theme, sorted by median. Orange = negative median, green = positive.
Box width $\propto$ mean comments/post; wider boxes indicate
higher community engagement. The strong negative correlation between sentiment
and engagement ($\rho=$ $-0.72$, $p<0.001$) is visible:
the widest, most-engaged boxes cluster at the negative end. }
\label{fig:sentiment}
\end{figure}

\subsection{Sentiment and Community Engagement}
Figure~\ref{fig:sentiment} shows that sentiment and community
engagement are deeply intertwined: the two measures are strongly
negatively correlated across themes (Spearman $\rho=$ $-0.72$,
$p<0.001$), meaning the most negatively charged themes are also the
ones that most mobilise communities to reply.
The corpus overall skews negative---13 of 17 themes have a negative
median RoBERTa score, and the corpus-wide median is $-0.29$ (64.2\% negative, 29.0\% positive)---but the degree of negativity
predicts community bandwidth more strongly than theme size does.
We use mean comments per post as our engagement proxy
(Cmts/post in Table~\ref{tab:lda_topics}; corpus mean 16.2),
treating it as an inverse measure of debatability: posts that provoke
disagreement or personal stakes generate more replies than posts treated
as settled questions.

The highest-conflict cluster (Figure~\ref{fig:sentiment}, widest
boxes) exemplifies this dynamic.
\emph{Frontline AI Reactions} (median $-0.61$; 37.5 cmts/post) and
\emph{Misconduct Enforcement} ($-0.56$; 35.5) are not the two largest
themes by corpus share, yet they generate by far the most replies per
post.
\emph{Personal Misconduct Narratives} ($-0.45$; 24.9),
\emph{AI Writing \& Evasion} ($-0.21$; 21.9), and
\emph{Assessment Redesign} ($-0.14$; 22.7) round out the top
engagement cluster (all five $\geq$21 cmts/post): each involves
personal stakes, institutional conflict, or injustice, triggering
solidarity and advice-seeking responses.
Notably, \emph{AI Job Displacement} ( $-0.41$; 8.0 cmts/post, sharpened relative to the unfiltered corpus because the dropped records were tangential AI-degree posts) breaks the pattern:
despite being markedly negative, it generates little engagement.
The key distinction is immediacy, as abstract societal fears remain diffuse, while personal accusations or enforcement dilemmas demand a direct response.

At the opposite pole, the four themes with positive medians are also
the least engaging.
\emph{AI Tool Selection} ($+0.17$; 4.7 cmts/post), \emph{Degrees \&
Programs} ($+0.13$; 4.4) had the most posts yet
the lowest engagement in the corpus and \emph{Job Applications}
($+0.05$; 5.7) are solution and aspiration-oriented, resulting in constructive responses but shallow
threads.
\emph{AI-Assisted Workflow} ($+0.14$; 10.6) is a partial exception;
its help-seeking character draws moderate practical engagement without the emotional intensity of conflict themes.


\FloatBarrier
\section{RQ3: Teacher-Student Co-Discussion}
\label{sec:teacher-student}

\begin{figure}[!t]
\centering
\includegraphics[width=\columnwidth,trim=0 0 0 30,clip]{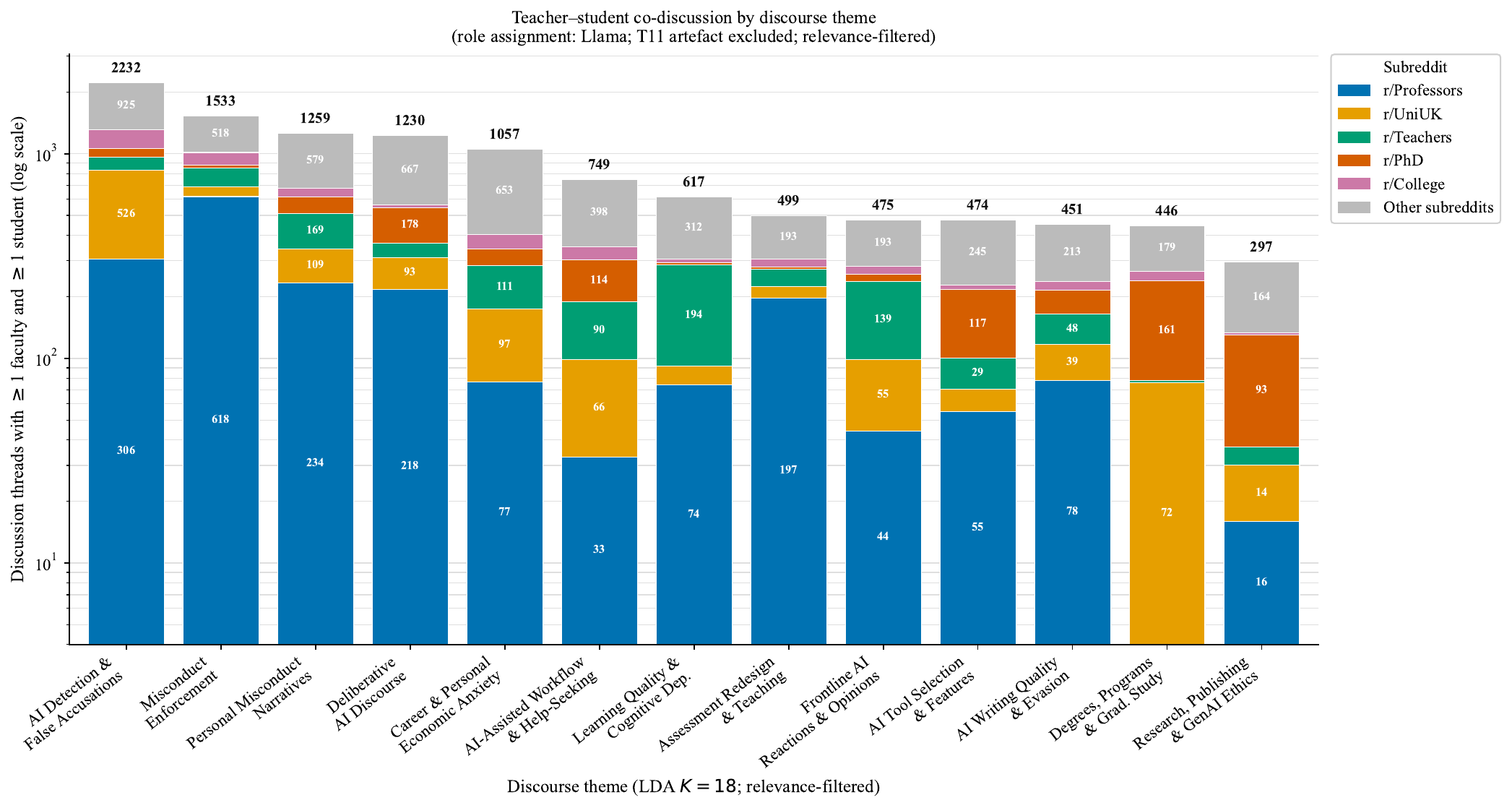}
\caption{Threads containing $\geq$1 faculty and $\geq$1 student participant,
by discourse theme (LDA $K=18$) and subreddit (top 5 shown; remainder as ``Other'').
Bars sorted by total count (descending); topics with $<$90 mixed threads omitted.}
\label{fig:teacher_student_topics}
\end{figure}

We now turn to the subset of threads where faculty and students participate
together, asking which discourse themes and communities hosting them most reliably draw cross-role
engagement.
A thread containing at least one inferred faculty author and one inferred student author is \emph{mixed}. 11,804 of 67,682 threads (17.4\%) are mixed.


Figure~\ref{fig:teacher_student_topics} shows the count of mixed threads by theme, coloured by the top five subreddits globally. The dominant theme for cross-role contact is \emph{AI Detection \&
False Accusations} (2,228 mixed threads, 18.9\% of all mixed threads).
This is the topic concerned with AI-writing detection, false
AI accusations, and evidence used in related disciplinary
proceedings. AI detection disputes involve an accuser (faculty) and a defendant
(student), making this a context where cross-role participation
is almost mandatory. Notably, r/UniUK contributes the
largest single-subreddit share (528 threads), exceeding r/Professors
(307) even though r/UniUK allocates 24.0\% of its AI discourse to
AI Detection against r/Professors's 9.7\%, suggesting that AI detection
is a particularly visible concern in UK higher-education communities.

The second-ranked theme, \emph{Misconduct Enforcement} (1,587 threads), is
closely related: it covers institutional responses, grade penalties, and
appeals. Here r/Professors accounts for 635 threads (40.0\%), by far the
largest subreddit share of any theme, showing that faculty are
active participants when discussing institutional sanctions.
Together, AI detection and enforcement account for 32.3\% of all
mixed threads, confirming that academic integrity conflicts are a primary
site of teacher-student co-discussions.

\emph{Deliberative AI Discourse} (1,244 threads) is the third-ranked theme and
represents more theoretical co-discussion: these threads contain
conversations about the legitimacy, risks, and future of
AI in education with participants from both instructor and student roles. The subreddit distribution is fairly balanced:
r/Professors (221), r/PhD (184), and r/AskAcademia (128) contribute the most,
suggesting that faculty-facing, graduate-research, and mixed-audience
communities are primary venues for intellectual AI deliberation.

Below the three, mixed-thread counts taper off. \emph{Career \&
Personal Economic Anxiety} (1,137), \emph{AI-Assisted Workflow \& Help-Seeking}
(769), and \emph{Learning Quality \& Cognitive Dependency} (624) attract moderate cross-role engagement.

\emph{Learning Quality \& Cognitive Dependency} stands out with a majority of secondary and general education communities.
r/Teachers accounts for 197 of 624 threads (31.6\%), the highest
share across all themes, while r/education (106, 17.0\%) and r/edtech
(72, 11.5\%) together contribute a further 28.5\%, confirming that concerns
about AI eroding learning and cognitive development are concentrated in K-12
and edtech communities.
The near-absence of r/PhD and r/UniUK suggests concerns
of cognitive depth come from different stakeholders than the academic-integrity conflicts dominating
cross-role discussion in HE-facing communities.

\noindent\textbf{Community Patterns:} Across all themes, r/Professors, r/UniUK, and r/Teachers are the three
largest hosts of mixed-participation threads (2,013; 1,246; 1,218 threads
respectively). r/Professors topping the list is meaningful:
though a faculty-facing community by name, its subject matter
(student behaviour, classroom dynamics, institutional policy) reliably draws
student voices into the fold.
r/UniUK ranks second overall, driven largely by its concentration in the
AI Detection theme. r/Teachers rises to third, reflecting
that the cross-role contacts generated by K-12 teaching.

The distribution is strongly right-skewed by theme. r/Professors dominates
\emph{Misconduct Enforcement} (635 of 1,587; 40.0\%) while
contributing proportionally less to \emph{Deliberative AI Discourse} (221 of 1,244; 17.8\%).
Conversely, r/PhD contributes disproportionately to deliberative threads
(184; 14.8\%) relative to its presence in enforcement discussions,
pointing to a participation norm difference: faculty-facing communities host
enforcement discussions, while graduate-research spaces host substantive debate.

\noindent\textbf{Who Initiates Cross-Role Discussion?:} For 6,527 mixed threads (55\%) we have a relevant root record that has a poster classified as a \textsc{Faculty} or a \textsc{Student}. The dominant initiation role falls to students at 68.4\% (4,466 threads) compared to faculty's 31.6\% (2,061 threads), but patterns vary substantially by discourse theme; per-theme percentages below are computed on each theme's binary-rooted subset so student and faculty shares sum to 100\%.

Two themes are strongly student-initiated: \emph{AI Detection \& False
Accusations} (79.2\% student / 20.8\% faculty; $n{=}1{,}536$ binary-rooted
mixed threads) and \emph{Career \& Personal Economic Anxiety}
(83.3\% student / 16.7\% faculty; $n{=}461$). In both, the
primary concern originates from the student perspective, either defending
against an accusation or anxieties about future employment. Faculty initiation, by contrast, peaks on themes where their professional
role grants framing authority: \emph{Assessment Redesign \& Teaching} is the
only majority-faculty theme (56.7\% faculty / 43.3\% student; $n{=}254$),
and faculty initiate roughly half of \emph{Misconduct Enforcement}
(48.8\% faculty / 51.2\% student; $n{=}893$) and
\emph{Learning Quality \& Cognitive Dependency}
(48.3\% faculty / 51.7\% student; $n{=}344$) threads, bringing
classroom-level enforcement, observed cognitive impact, and
the design of AI-resistant assessments into cross-role discussion. 


Reply directionality shows the complementary face of the initiation asymmetry: because students start most cross-role threads, faculty produce most cross-role replies. Among the 71,249 reply edges in mixed threads where both
authors carry a binary role label, cross-role edges account for 40.7\%
(28,967 edges); faculty replying to students substantially outnumber the reverse: 17,904 vs.\ 11,063 edges. The dominant pattern is student initiated threads with responses from multiple faculty members, each replying once. This corpus-wide direction is community-conditioned: in student-leaning subreddits (r/UniUK, r/PhD, r/CollegeRant, r/AskProfessors) faculty reply to students 2--3$\times$ as often as the reverse, while in r/Professors and r/Teachers the pattern inverts, hosting communities.
Genuinely bidirectional exchange, threads with at least one cross-role reply in
\emph{both} directions makes up only 22.9\% of mixed
threads.

\noindent\textbf{Thread Engagement and Structural Depth:} We compare mixed threads against same-role threads with at least two records to ensure a more reliable analysis. Under this matched-floor restriction (mixed $n{=}11{,}804$; faculty-only
$n{=}9{,}181$; student-only $n{=}5{,}546$), mixed threads remain
substantially more engaged than same-role threads across all structural
metrics. Mixed threads have a median of 6~records per thread vs.\ 3~records for faculty-only and 2~records for student-only threads
(Mann-Whitney $p < 0.001$ for both comparisons). Median
conversation depth is 2 comment levels for mixed threads vs.\ 1 for
same-role threads (both $p < 0.001$). Median \emph{thread lifespan}
(time between a thread's earliest and latest records) is 20.0~hours for mixed
threads vs.\ 9.5~hours for faculty-only and 4.9~hours for student-only
threads ($p < 0.001$ for both), indicating that cross-role participation
is associated with lengthier conversations.

\subsection{Sentiment Patterns Within Mixed Threads}

\noindent\textbf{Theme-level sentiment and cross-role contact:} The themes that attract the most cross-role contact are also the most negatively charged.
Among the 13 themes with $\geq$90 mixed threads, the
per-theme mixed-thread count is negatively correlated with median sentiment (Spearman$\rho{=}-0.47$, $p{=}0.103$, $N{=}13$);
the full-corpus correlation falls short of significance because of
a single structural outlier: \emph{Frontline AI Reactions} (T16) is the
most negative theme (median $-0.61$) yet generates among the fewest
mixed threads (485), because it is concentrated in communities
(r/Teachers, r/CollegeRant) where faculty and their students rarely
share the same Reddit spaces.
Restricting to the 12 HE-facing themes, the correlation strengthens to
$\rho{=}-0.72$ ($p{=}0.008$), confirming that within higher education, negativity and cross-role contact are systematically coupled.

\emph{AI Detection \& False Accusations} (2,228 mixed threads; corpus
median $-0.41$) and \emph{Misconduct Enforcement} (1,587; $-0.56$) dominate cross-role contact because their adversarial structure necessarily brings together accusing and accused parties. At the other end, themes with positive sentiment attract far fewer
mixed threads: \emph{AI-Assisted Workflow} (769; $+0.14$),
\emph{AI Tool Selection} (490; $+0.17$), and \emph{Degrees \&
Programs} (513; $+0.13$) are communities where posts get answered
with little cross-role debate.


\noindent\textbf{Role-level sentiment:} Both roles post with negative median sentiment in mixed threads,
with students less negative, although the difference is negligible ($-0.369$ vs.\ $-0.373$;
Mann-Whitney $p = 0.04$).
Both roles are predominantly negative (faculty 70.9\%, students 69.1\%),
with students somewhat more likely to post positively (24.7\% vs.\ 22.5\%).

The role gap's direction varies by theme.
In \emph{AI Detection \& False Accusations} threads, faculty are less negative
than students (median $-0.431$ vs.\ $-0.467$, $p<0.001$), consistent with
students posting about accusations skew more distressed than their faculty counterparts.
In \emph{Degrees, Programs \& Graduate Study} threads, students are less
negative than faculty (median $+0.071$ vs.\ $-0.009$, $\Delta=-0.080$, $p<0.001$), reflecting that students frame degree decisions more
optimistically than faculty.
The largest cross-role gaps are in \emph{AI-Assisted Workflow} threads
(student median $+0.016$ vs.\ faculty $-0.117$, $\Delta=-0.133$, $p<0.001$): students frame
help-seeking interactions positively while faculty engage with more critically.
For most remaining themes---\emph{Deliberative AI Discourse}, \emph{Career \&
Personal Economic Anxiety}, and \emph{AI Writing Quality \& Evasion} the
median differences are small ($|\Delta|<0.05$) and not statistically significant,
indicating broadly parallel negative sentiment across roles.

\noindent\textbf{Initiator role and sentiment:}
Threads initiated by faculty are less negative overall than
student-initiated threads (median $-0.351$ vs $-0.395$, $p<0.001$).
Faculty post less negatively when responding in faculty-initiated threads than in student-initiated ones (median $-0.347$ vs $-0.411$, $p<0.001$),
suggesting faculty who raise issues frame them constructively but reply with emotional reactions. Students do not show an analogous pattern: their cross-initiator gap (medians $-0.369$ vs.\ $-0.381$) is too small to be substantively meaningful, so this pattern is faculty-specific.

\noindent\textbf{Escalation Analysis:}
We measure sentiment trajectory as the OLS slope of sentiment score
against comment depth within threads with at least four records.
In mixed threads, the trajectory differs sharply by topic.
\emph{Misconduct Enforcement} shows the most consistently improving
trajectory (median $+0.030$; 53.1\% of threads improving), showing
deeper conversation within enforcement-related threads tends
toward resolution or less heated exchange. This pattern is highly
robust: requiring threads to have at least 10, 20, or 30 records yields
essentially the same median ($+0.026$ to $+0.029$) and improving
proportion ($\approx 52\text{-}53\%$). \emph{AI Detection} is the most
balanced (49.5\% improving, 47.6\% deteriorating) at every examined threshold. \emph{AI Tool Selection} (61.4\% deteriorating) and
\emph{Degrees \& Graduate Study} (62.1\% deteriorating) show the
strongest negative trajectories at baseline threshold; both patterns
remain in the same direction for threads up to 10 records, after
which sample sizes become too small to read confidently. In conclusion, topics shape trajectory substantially: some discussions reliably de-escalate with depth, while others tend to deteriorate. 

\FloatBarrier

\section{Discussion \& Conclusion}

The discourse is dominated by academic-integrity enforcement, AI-detection disputes, cognitive dependency, and uncertainty about the future value of degrees. Collectively, these themes indicate that the rapid adoption of GenAI has generated a persistent governance challenge for education.

The clearest evidence comes from the academic-integrity cluster, which accounts for over one-third of AI-related discourse. \emph{Misconduct Enforcement, AI Detection and False Accusations, Personal Misconduct Narratives}, and \emph{AI Writing Quality and Evasion} together depict an adversarial ecosystem in which students, faculty, and institutions operate under mutual distrust. High engagement further suggests integrity disputes consume substantial community attention. This extends prior Reddit-based work~\cite{wu2024reacting,devito2025unpacking}, showing such concerns have not faded since the launch period but are an enduring feature of educational AI discourse. Crucially, four themes are absent from earlier taxonomies: \emph{Career \& Personal Economic Anxiety} (10.0\%), \emph{AI Writing Quality \& Evasion} (6.4\%), \emph{Research \& Publishing Ethics} (2.1\%), and \emph{Healthcare \& Medical Education} (0.4\%) make up 19\% of the corpus. These reflect concerns that emerged or grew substantially as AI adoption matured beyond earlier observation windows.

Our findings also expose gaps between institutional policy and stakeholder experience. Although universities have increasingly formalised AI policies, Reddit discussions reveal continued uncertainty about evidentiary standards, detector reliability, faculty workload, and procedural fairness. Students describe anxiety, reputational harm, and perceived injustice from false-positive accusations, while faculty report issues balancing teaching responsibility with limited guidance and unreliable tools. Mixed-role threads are particularly adversarial, suggesting cross-role engagement arises not through collaborative deliberation but through conflict between accuser and accused, instructor and student.

However, the discourse is not solely conflict-oriented. Over time, the discussion has shifted from early-detection-focused panic to more pragmatic questions about the pedagogical effects of AI and its role in shaping career opportunities. For instance, K--12 teacher communities emphasise learning quality and cognitive dependence, whereas professional programmes such as nursing, medicine, and law express heightened anxiety about implications for future employability, professional competence, and credential values. This suggests that educational communities are moving from debates about \emph{permitting} to \emph{integration} of AI.

Such pragmatic concerns also highlight unresolved questions about teacher workload, student learning, and the legitimacy and equity of AI-enforcement practices, all of which remain significant sources of tension for students and teachers and need consideration within wider educational policies. Our findings point to a need to move beyond detection-centred compliance and allow teachers and students to negotiate responsible, transparent, and meaningful AI use. Restorative alternatives, including co-designed assessments, student-staff policy forums, transparent disclosure norms, and AI-literacy support, can reframe cross-role contact around GenAI as deliberative rather than accusatory. Our findings support a governance model in which faculty and students engage with GenAI through shared responsibility, clear pedagogy, and institutional trust, rather than relying on the current detection-focused systems.

\noindent\textbf{Limitations:} Our corpus skews male, North American, 
and Anglophone, excluding non-English platforms and private institutional 
channels (Slack, LMS forums). 
\texttt{twitter-roberta-base-sentiment-latest}'s Twitter pre-training may 
misclassify Reddit-specific conventions. The Llama~3.3~70B role classifier 
covers $75.1\%$ of authors; in clinical communities 
(\textit{r/nursing}, \textit{r/medicalschool}, \textit{r/DentalSchool}), 
preceptors may be classified as \textsc{Faculty}, so Fac\% figures are 
upper bounds.

\bibliography{paper}
\section*{Ethics Statement}
This study received ethical approval from a relevant institutional board. All data used in this study was publicly posted on Reddit.
No personally identifiable information was collected or retained; deleted posts were excluded.

The aggregate findings, in particular which subreddits host the most
adversarial AI-integrity contact, and which themes most reliably draw
faculty-student conflict, could in principle be misused for
community targeting (astroturfing, harassment) or to market detection
tools whose unreliability we explicitly document. We mitigate this by
reporting at aggregate level, replacing all usernames with consistent SHA-256 hashes before release, and by framing the Discussion around restorative pedagogical redesign and
procedural fairness rather than detection or enforcement tooling. The
LLM relevance- and role-classification prompts are released for
transparency and reproducibility; they should not be repurposed for
identifying individual Reddit users.

\section*{Ethics Checklist}

\begin{enumerate}

\item For most authors...
\begin{enumerate}
    \item  Would answering this research question advance science without violating social contracts, such as violating privacy norms, perpetuating unfair profiling, exacerbating the socio-economic divide, or implying disrespect to societies or cultures?
    \answerYes{Yes, the corpus is built only from publicly posted Reddit content; no personally identifiable information is collected or retained; deleted posts are excluded; institutional ethics approval was obtained (see Ethical Statement). Findings are reported at the theme, subreddit, and role-aggregate level, never the individual level except for a few top posts that are paraphrased if they contain sensitive information. The work is aimed at informing governance and pedagogical design rather than profiling individuals.}
  \item Do your main claims in the abstract and introduction accurately reflect the paper's contributions and scope?
    \answerYes{Yes, the Abstract and Introduction (RQ1--RQ3) match the analyses reported in the Results, Teacher--Student, and Discussion sections: 270{,}929 records from 26 education subreddits, 17 themes obtained via LDA, sentiment--engagement correlations, and faculty--student co-discussion analysis.}
   \item Do you clarify how the proposed methodological approach is appropriate for the claims made?
    \answerYes{Yes, see Methods. Topic modelling choices (LDA $K{=}18$ selected by TC$\times$TD, BERTopic as convergent validation), sentiment scoring (twitter-roberta-base-sentiment-latest), and LLM-based role inference (Llama 3.3 70B with human validation) are each justified, validated, and tied to the specific RQ they serve.}
   \item Do you clarify what are possible artifacts in the data used, given population-specific distributions?
    \answerYes{Yes, the Limitations section flags that the corpus skews male, North American, and Anglophone, excludes non-English platforms and private institutional channels, that twitter-roberta sentiment may misread Reddit conventions, and that in clinical subreddits the role classifier may label preceptors as Faculty (so Fac\% figures are upper bounds). The Methods section also documents an r/HomeworkHelp spillover topic (T11) and an r/LawSchool ``LL.M.'' confound (T15/T04) that are filtered.}
  \item Did you describe the limitations of your work?
    \answerYes{Yes, see the Limitations section.}
  \item Did you discuss any potential negative societal impacts of your work?
    \answerYes{Yes, the Discussion engages explicitly with the adversarial dynamics, false-positive accusations, detector unreliability, and faculty--student distrust surfaced by the corpus, and argues for restorative redesign of cross-role contact rather than detection-centred compliance.}
      \item Did you discuss any potential misuse of your work?
    \answerYes{Yes, see the ``Potential misuse'' paragraph of the Ethical Statement: we identify community targeting and the marketing of unreliable detection tools as the principal misuse risks, and state the mitigations (aggregate-only reporting, no per-user or per-thread artefacts, framing of the Discussion around restorative redesign).}
    \item Did you describe steps taken to prevent or mitigate potential negative outcomes of the research, such as data and model documentation, data anonymization, responsible release, access control, and the reproducibility of findings?
    \answerYes{Yes, see Ethical Statement (public-only data, no PII retained, deleted posts excluded, institutional ethics approval), Methods (fixed random seed 42, documented hyperparameters, LLM prompts in Appendix)}
  \item Have you read the ethics review guidelines and ensured that your paper conforms to them?
    \answerYes{Yes.}
\end{enumerate}

\item Additionally, if your study involves hypotheses testing...
\begin{enumerate}
  \item Did you clearly state the assumptions underlying all theoretical results?
    \answerNA{NA, the paper is an empirical, observational analysis of public Reddit discourse and does not advance formal theoretical results requiring stated assumptions.}
  \item Have you provided justifications for all theoretical results?
    \answerNA{NA, no theoretical results are presented.}
  \item Did you discuss competing hypotheses or theories that might challenge or complement your theoretical results?
    \answerNA{NA, no theoretical results are presented; alternative interpretations of empirical findings are discussed in the Discussion section.}
  \item Have you considered alternative mechanisms or explanations that might account for the same outcomes observed in your study?
    \answerNA{NA, no theoretical results are presented.}
  \item Did you address potential biases or limitations in your theoretical framework?
    \answerNA{NA, no theoretical framework is proposed; empirical biases are addressed in the Limitations section.}
  \item Have you related your theoretical results to the existing literature in social science?
    \answerNA{NA, no theoretical results are presented; empirical findings are related to prior Reddit-based work in the Related Work and Discussion sections.}
  \item Did you discuss the implications of your theoretical results for policy, practice, or further research in the social science domain?
    \answerNA{NA, no theoretical results are presented; policy and pedagogical implications of the empirical findings are discussed in the Discussion section.}
\end{enumerate}

\item Additionally, if you are including theoretical proofs...
\begin{enumerate}
  \item Did you state the full set of assumptions of all theoretical results?
    \answerNA{NA, the paper contains no theoretical proofs.}
	\item Did you include complete proofs of all theoretical results?
    \answerNA{NA, the paper contains no theoretical proofs.}
\end{enumerate}

\item Additionally, if you ran machine learning experiments...
\begin{enumerate}
  \item Did you include the code, data, and instructions needed to reproduce the main experimental results (either in the supplemental material or as a URL)?
    \answerYes{Yes, see the Code and Data Availability statement in the Acknowledgements and Appendix~\ref{app:licences}: the analysis code is released for review at an anonymised mirror and will be de-anonymised on publication; the full AI-related corpus (270,929 records) is released with all user identifiers replaced by consistent SHA-256 hashes, together with per-record topic assignments, sentiment scores, and role classifications. Full hyperparameters, seeds, LLM prompts, and validation procedures are reported in Methods and Appendices~\ref{app:bertopic},~\ref{app:roledetection},~\ref{app:opus_b8}.}
  \item Did you specify all the training details (e.g., data splits, hyperparameters, how they were chosen)?
    \answerYes{Yes, see Methods and Appendix: preprocessing rules, LDA settings ($\alpha{=}\eta{=}1/K$, 10 passes, seed 42, vocabulary filters $\text{min\_df}{=}5$, $\text{max\_df}{=}0.90$), the coarse-and-fine $K$ sweep ($K{\in}\{5,\ldots,60\}$ then $\{11,\ldots,19\}$), BERTopic configuration (all-MiniLM-L6-v2, UMAP $n_\text{neighbors}{=}15$, $d{=}5$, HDBSCAN \texttt{min\_cluster\_size}{=}541, KeyBERT/MMR/POS refinement), and the Llama 3.3 70B role-classification and relevance prompts are fully reported.}
     \item Did you report error bars (e.g., with respect to the random seed after running experiments multiple times)?
    \answerYes{Yes, a six-seed LDA stability analysis confirms the $K{=}18$ selection (Appendix), inter-annotator and LDA-vs-human agreement are reported with Cohen's $\kappa$ at both topic and five-cluster levels, the Llama relevance classifier is benchmarked against two annotators with $\kappa$ and raw agreement, and quantitative claims are accompanied by effect sizes (median deltas) alongside $p$-values.}
	\item Did you include the total amount of compute and the type of resources used (e.g., type of GPUs, internal cluster, or cloud provider)?
    \answerYes{Yes, see Methods (Reproducibility and Compute): LDA on CPU ($\sim$2.3 h); BERTopic and RoBERTa sentiment scoring on a local NVIDIA RTX 4060 ($\sim$35 min and $\sim$1 h); Llama 3.3 70B relevance and role classification via an institutional inference endpoint ($\sim$39 h batched API calls).}
     \item Do you justify how the proposed evaluation is sufficient and appropriate to the claims made?
    \answerYes{Yes, model selection uses Topic Quality (TC$\times$TD) with both a coarse and a fine $K$-sweep; BERTopic provides convergent validation through a separate density-based pipeline; two expert annotators provide a 583-post human validation of topic assignment; the Llama relevance classifier is validated against the same human-coded sample and exceeds the human inter-annotator ceiling.}
     \item Do you discuss what is ``the cost`` of misclassification and fault (in)tolerance?
    \answerYes{Yes, the Methods section explicitly anchors paper-level interpretation at the five-cluster level because topic-level boundary disagreements (e.g.\ inside the Academic Integrity ``cheating triangle'') drive the moderate topic-level $\kappa$; the Limitations section bounds the role classifier's Fac\% figures in clinical subreddits as upper bounds; and the relevance-filter step removes an entire incoherent topic (T11) and a substantial irrelevant share of T15.}

\end{enumerate}

\item Additionally, if you are using existing assets (e.g., code, data, models) or curating/releasing new assets, \textbf{without compromising anonymity}...
\begin{enumerate}
  \item If your work uses existing assets, did you cite the creators?
    \answerYes{Yes, all third-party assets are cited if they are used in the main methodology: the Arctic Shift Reddit API, gensim LDA, BERTopic, all-MiniLM-L6-v2, twitter-roberta-base-sentiment-latest, NLTK WordNet lemmatisation, and Llama 3.3 70B Instruct.}
  \item Did you mention the license of the assets?
    \answerYes{Yes, see Appendix~\ref{app:licences} (``Software and Model Licences''): a table enumerates the licences of every third-party library, pre-trained model, and external API used, alongside the terms governing the Reddit source data.}
  \item Did you include any new assets in the supplemental material or as a URL?
    \answerYes{Yes, the analysis pipeline (filters, LDA/BERTopic harness, sentiment scoring, role-classification harness, figure-generation scripts) is released at an anonymised mirror linked from the Code and Data Availability statement in the Acknowledgements, together with the full anonymised corpus (user identifiers replaced by SHA-256 hashes) and Arctic Shift collection scripts.}
  \item Did you discuss whether and how consent was obtained from people whose data you're using/curating?
    \answerYes{Yes, see the ``Consent and minimisation'' paragraph of the Ethical Statement: the institutional ethics review treated the corpus as public, low-risk, observational data; individual consent was not sought, and the mitigations (Arctic Shift API under non-commercial terms, exclusion of deleted posts, no retention of usernames/IDs, aggregate-only reporting, paraphrased quotations) are explicitly described.}
  \item Did you discuss whether the data you are using/curating contains personally identifiable information or offensive content?
    \answerYes{Yes, the Ethical Statement states that no personally identifiable information was collected or retained and that deleted posts were excluded. Reddit content can contain coarse or offensive language; we therefore report and interpret results at the theme, subreddit, and role-aggregate level rather than at the individual user or post level.}
\item If you are curating or releasing new datasets, did you discuss how you intend to make your datasets FAIR (see FORCE11 (2020))? 
\answerYes{Yes. The AI-education corpus (270,929 records) will be deposited with a persistent DOI on publication. Each record includes the post/comment ID, a consistent SHA-256 hashed user identifier, subreddit, timestamp, text, and derived fields (topic assignment, sentiment score, role classification). The dataset is Findable (persistent DOI), Accessible (open download), Interoperable (JSON Lines with a documented schema), and Reusable (CC BY 4.0 licence with full provenance documented in Methods and Appendix~\ref{app:licences}).}
\item If you are curating or releasing new datasets, did you create a Datasheet for the Dataset (see Gebru et al. (2021))? 
\answerYes{Yes. Dataset documentation covering motivation, composition, collection process, preprocessing, intended uses, distribution, and maintenance is provided across the Methods section and Appendix~\ref{app:licences}. User identifiers are replaced by SHA-256 hashes; deleted posts are excluded; text is released as collected from the Arctic Shift API.}
\end{enumerate}

\item Additionally, if you used crowdsourcing or conducted research with human subjects, \textbf{without compromising anonymity}...
\begin{enumerate}
  \item Did you include the full text of instructions given to participants and screenshots?
    \answerNA{NA, no crowdworkers or interactive human participants were recruited; annotation was performed by two expert annotators on the research team using an internal codebook. The full Llama 3.3 70B system prompts used for relevance and role classification are reported in the Appendix.}
  \item Did you describe any potential participant risks, with mentions of Institutional Review Board (IRB) approvals?
    \answerYes{Yes, the Ethical Statement notes that the study received ethical approval from a relevant institutional board; risk to data subjects is minimised through public-only data, exclusion of deleted posts, no retention of personally identifiable information, and aggregate-only reporting.}
  \item Did you include the estimated hourly wage paid to participants and the total amount spent on participant compensation?
    \answerNA{NA, no paid participants or crowdworkers were used.}
   \item Did you discuss how data is stored, shared, and deidentified?
   \answerYes{Yes, the Ethical Statement specifies that no personally identifiable information was collected or retained and that deleted posts were excluded; the full AI-related corpus is released with all user identifiers replaced by consistent SHA-256 hashes, and reported analyses are aggregated at the theme/subreddit/role level.}
\end{enumerate}

\end{enumerate}

\FloatBarrier
\clearpage
\appendix
\section{Corpus Construction \& Keyword Filter}
\label{app:patcounts}\label{app:keywords}\label{app:validation}


\begin{table}[h]
\centering
\small
\setlength{\tabcolsep}{5pt}
\begin{tabular}{lrrr}
\toprule
\textbf{Keyword / filter} & \textbf{Posts} & \textbf{Cmts} & \textbf{Total} \\
\midrule
\multicolumn{4}{l}{\textit{Corpus size (bare-AI \textbf{or} named-tool match)}} \\
\quad Union filter & 36,437 & 234,492 & 270,929 \\
\midrule
\multicolumn{4}{l}{\textit{Named-tool mentions in union corpus (non-mutually exclusive)}} \\
ChatGPT / GPT variants        &  3,858 &  16,304 & 20,162 \\
Turnitin                       &    953 &   3,867 &  4,820 \\
Grammarly                      &    453 &   3,425 &  3,878 \\
LLM                            &    409 &   2,547 &  2,956 \\
GPTZero / ZeroGPT / WinstonAI &    275 &   1,214 &  1,489 \\
Gemini                         &    317 &     819 &  1,136 \\
Claude                         &    295 &     803 &  1,098 \\
OpenAI                         &    236 &     824 &  1,060 \\
Copilot                        &    115 &     513 &    628 \\
Perplexity                     &    228 &     370 &    598 \\
GenAI (fused)                  &    120 &     453 &    573 \\
Quillbot                       &    117 &     359 &    476 \\
NotebookLM                     &     76 &     252 &    328 \\
\bottomrule
\end{tabular}
\caption{Corpus size and named-tool mention counts.
The 270,929 records are the union of bare-AI
(\texttt{\textbackslash b(AI|ai)\textbackslash b}) and named-tool matches;
62,468 records (23.1\%) match named tools only and would be absent from a
bare-AI-only corpus.
Named-tool counts are drawn from the full union corpus and are
non-mutually exclusive.}
\label{tab:patcounts}
\end{table}

\subsection*{Removed Keywords}

Three model-name keywords were removed due to lexical ambiguity and negligible unique signal:
\texttt{bard} (645 total mentions, 434 unique-only; rebranded to Gemini Feb~2024),
\texttt{llama} (345 total, 276 unique-only; primarily the animal name or developer-facing Meta LLaMA with minimal classroom uptake),
and \texttt{mistral} (43 total, 18 unique-only; Mediterranean wind / French surname).
By contrast, \texttt{perplexity} (1,022 total; 542 unique-only) was retained as a
widely used student-facing research tool with lower ambiguity in educational contexts.

\section{BERTopic: Configuration, Natural Cluster Discovery, and Convergent Validation}
\label{app:bertopic}

\subsection*{Configuration}

BERTopic \cite{grootendorst2022bertopic} was run with \texttt{all-MiniLM-L6-v2}
sentence embeddings (GPU-accelerated, 384 dimensions), UMAP dimensionality
reduction ($n_\text{neighbors}=15$, $d=5$, cosine metric, \texttt{random\_state}=42),
and HDBSCAN density-based clustering
(\texttt{min\_cluster\_size}=$\max(50,\lfloor N/500\rfloor)=541$,
Euclidean metric, EOM selection).
Topic representations were refined using a three-stage pipeline:
KeyBERT-inspired relevance re-ranking,
Maximal Marginal Relevance (MMR, $\lambda=0.3$),
and spaCy part-of-speech filtering retaining nouns, adjectives, and proper nouns.
We set \texttt{top\_n\_words}=50 to give the refinement chain sufficient
candidate vocabulary before narrowing to the final 25 words per topic.
We swept $K \in \{10,11,12,13,14,15\}$ as a focused range around the LDA
optimum; $C_v$ coherence was evaluated against raw-text tokenisations
(not lemmatised forms) to match the vocabulary produced by BERTopic's c-TF-IDF
representation.

\subsection*{Natural Cluster Discovery}

A key observation from BERTopic is that HDBSCAN's density-based clustering
converges to a small number of natural clusters---well below the requested
$nr\_topics$---regardless of how large $K$ is set:

\begin{table}[h]
\centering
\small
\begin{tabular}{rrrrrr}
\toprule
\textbf{Req.\ $K$} & \textbf{Natural} & \textbf{Effective} & \textbf{TC} & \textbf{TD} & \textbf{Outliers} \\
\midrule
10 &  6 &  5 & 0.544 & 0.911 &  3,956 \\
11 &  6 &  5 & 0.544 & 0.911 &  3,956 \\
12 & 12 & 11 & 0.519 & 0.902 & 14,015 \\
13 &  6 &  5 & 0.544 & 0.911 &  3,956 \\
14 &  6 &  5 & 0.544 & 0.911 &  3,956 \\
15 &  6 &  5 & 0.544 & 0.911 &  3,956 \\
\bottomrule
\end{tabular}
\caption{BERTopic results for requested $K \in \{10,\ldots,15\}$.
Natural = HDBSCAN cluster count before reduction;
Effective = non-outlier topics after representation refinement;
TC = $C_v$ coherence on full 270,929-document raw-text
tokenisation (not directly comparable to LDA TC, which uses preprocessed tokens);
Outliers = documents assigned to topic $-1$.
K=10/11/13--15 produce identical models: HDBSCAN finds exactly 6
density peaks every run, yielding 5 non-outlier topics.}
\label{tab:bertopic_clusters}
\end{table}

This saturation at 5--11 effective topics reflects the inherent density
structure of the 270,926-document corpus in UMAP-reduced space under
\texttt{min\_cluster\_size}=541.
For five of the six runs (K=10/11/13/14/15), the results are bitwise
identical---same clusters, same coherence, same outlier count---confirming
that HDBSCAN found a single stable solution regardless of the requested $K$.
Only K=12 finds a second stable state with 12 natural clusters, producing
a different (lower-coherence, higher-outlier) partition.
This is a known property of HDBSCAN-based topic models at scale: with a large
corpus and a minimum cluster size proportional to corpus size, the algorithm
identifies only the most prominent density peaks.

\subsection*{Implications for Validation}

HDBSCAN consistently recovers only 5 effective topics from this corpus,
well below the LDA optimum of $K=18$.
BERTopic therefore cannot serve as an independent $K$-selection tool here.
Its value is qualitative: the 5 stable density peaks correspond to
broad macro-themes (AI detection/integrity, learning quality, practical AI
use, career/professional impact, and general discourse), confirming that
these are the most strongly signal-dense clusters in the embedding space.
All five align with major themes in the LDA taxonomy, providing convergent
validation of the most prominent discourse dimensions.
LDA's finer $K=18$ partition recovers additional coherent sub-themes
(assessment redesign, institutional policy, research ethics) that BERTopic
subsumes into broader clusters or assigns to outliers.
Themes corroborated by both models are treated as more robust in
Section~\ref{sec:results}.

\subsection*{BERTopic 11-Topic Model vs.\ LDA 18-Topic Model}

To examine topic-level correspondence, we compare the top-15 words of each
BERTopic topic (from the K=12 run, which is the only run that produces an
11-topic partition) against the 18 LDA topics.
Table~\ref{tab:btlda_crosscheck} summarises the mapping; alignment strength
is assessed by keyword overlap.

\begin{table*}[t]
\centering
\small
\resizebox{\textwidth}{!}{%
\begin{tabular}{clp{4.2cm}p{3.8cm}l}
\toprule
\textbf{BT} & \textbf{BT signal words} & \textbf{LDA match} & \textbf{Notes} & \textbf{Align} \\
\midrule
BT00 & law, legal, lawyers, case  & L04 (law, legal, policy)         & LDA conflates law with broader institutional policy & Partial \\
BT01 & llm, law, tax, bar, llb    & L09 + L04                        & BERTopic isolates law-school LLM-degree discussions (LLM = \emph{Legum Magister}); LDA does not separate this from degree/program discourse & Partial \\
BT02 & em, dashes, dash, punctuation & L02 (detection, turnitin)     & Very specific sub-theme: em-dash overuse as AI stylometric signal; LDA subsumes into general detection topic & Partial \\
BT03 & grammar, spelling, writing, check & L17 (writing, essay, word) + L02 & Writing tools and Grammarly cluster; LDA separates grammar assistance (L17) from detection (L02) & Strong \\
BT04 & gemini, claude, chatgpt, free, access & L03 (tool, free, app, platform) + L13 & AI tool comparison and selection & Strong \\
BT05 & medical, nursing, patients, medicine & L15 (patient, nurse, nursing, doctor) & Clean one-to-one match & Exact \\
BT06 & phd, research, university, degree & L09 (degree, research, experience) + L06 (paper, citation) & Graduate/research context; LDA finer---separates degree-seeking (L09) from academic publishing (L06) & Strong \\
BT07 & llm, llms, writing, students & L10 / L16 (generic normalisation) & Residual general LLM discussion; no single LDA match & Weak \\
BT08 & chatgpt, gpt, students, write & L05 (student, assignment, cheating) + L13 & General student ChatGPT use; LDA finer---separates misconduct (L05) from help-seeking (L13) & Strong \\
BT09 & students, work, class, paper & L05 / L08 (exam, assignment, course) & Generic student/classroom signal; LDA resolves into assessment and misconduct sub-themes & Strong \\
BT10 & email, letter, recommendation, professor & L14 (letter, email, application) & Clean one-to-one match & Exact \\
\bottomrule
\end{tabular}
}%
\caption{Mapping of BERTopic 11-topic model (K=12 run) to LDA $K{=}18$ topics
by top-word overlap. Alignment: \emph{Exact} = one-to-one semantic match;
\emph{Strong} = majority of signal words shared, LDA adds sub-theme resolution;
\emph{Partial} = overlap on primary domain but models diverge on framing;
\emph{Weak} = BERTopic topic too generic for a single LDA match.}
\label{tab:btlda_crosscheck}
\end{table*}

\paragraph{LDA topics with no BERTopic counterpart.}
Four LDA topics are not recovered by BERTopic at all.
\emph{L00} (said, got, asked, told, first-person narrative framing) captures
the personal-account stance that appears across misconduct and false-accusation
posts; HDBSCAN does not cluster on stylistic or narrative stance.
\emph{L07} (robot, replace, company, tech, society, machine) captures AI
job-displacement anxiety, which BERTopic disperses across BT07/BT08 as
background noise.
\emph{L12} (job, pay, year, going, life, better) captures career and
future-value discourse; similarly absorbed into generic BERTopic topics.
\emph{L11} is a mixed artefact topic in both models and is excluded from
the interpretive analysis (see Section~\ref{sec:results}).

\paragraph{Summary.}
Three exact or near-exact matches (BT05$\leftrightarrow$L15,
BT10$\leftrightarrow$L14, BT04$\leftrightarrow$L03) and five strong
alignments confirm that the most semantically dense discourse regions in the
corpus are robustly identified by both methods.
BERTopic additionally isolates one sub-niche absent from LDA: law-school
LLM-degree discussions (BT01), where ``LLM'' refers to the \emph{Legum
Magister} qualification rather than large language models---a disambiguation
LDA's bag-of-words representation cannot make.
Conversely, LDA recovers three discourse types BERTopic misses entirely:
personal-narrative framing (L00), AI job-displacement anxiety (L07), and
career/future-value concerns (L12).
These differences are consistent with the respective model designs: BERTopic
clusters on dense semantic neighbourhoods in embedding space, while LDA
recovers thematic co-occurrence patterns including discourse stance and
societal-level framing.
We therefore use LDA $K{=}18$ (seventeen distinct themes, T11 excluded as
artefact) as the primary analytical framework, with BERTopic convergence
as corroborating evidence for the highest-density themes.

FASTopic was also attempted but its dense
topic--document matrix exceeded the 8\,GB VRAM budget on our corpus
and was excluded.

\section{Token-Length Distribution and Minimum-Length Threshold}
\label{app:tokenhist}

Figure~\ref{fig:tokenhist} shows the distribution of record lengths (tokens
counted by whitespace splitting) across all 270,929 records in the broad
AI-filtered corpus.
The tail is capped at 300 tokens for legibility; roughly 5\% of records exceed
that length.
Vertical dashed lines mark four candidate minimum-length thresholds with the
corresponding share of the corpus that would be removed.

\begin{figure}[h]
\centering
\includegraphics[width=\columnwidth]{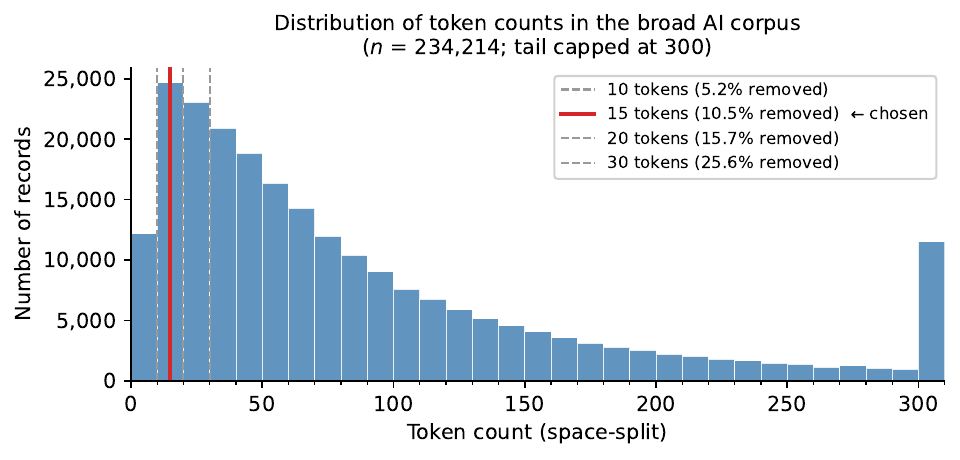}
\caption{Distribution of token counts (space-split) in the broad AI corpus
($n=270{,}929$). Dashed vertical lines show candidate minimum-length thresholds
and the fraction of records they would remove.
The chosen threshold of 15 tokens is highlighted.}
\label{fig:tokenhist}
\end{figure}

To choose the threshold we manually inspected a stratified random sample of
40 records in the 5--20 token range.
Three qualitative bands emerged:

\textbf{5--8 tokens (remove).}
This range is dominated by deleted-post stubs (titles whose body was
subsequently removed), bare hyperlinks to AI tools, and semantically empty
fragments (e.g., \emph{``it still sounds somewhat like ai''}). None carry
topical signal useful for modelling.

\textbf{9--14 tokens (borderline).}
Records in this range include some genuinely on-topic short comments
(e.g., \emph{``Why are you assigning homework that AI can do''},
\emph{``AI is great at writing these types of replies''}) alongside noise
such as link-only posts and single-word reactions.
The majority are complete syntactic units but lack the contextual detail
needed for reliable topic assignment.

\textbf{15--20 tokens (keep).}
Nearly all sampled records in this range express a clear, complete thought
with specific AI-in-education content (e.g., \emph{``Worried about
wrongfully accusing students of using AI, unsure how to deal with denial''},
\emph{``How do you know there aren't students using AI cleverly to boost
their grade without your notice?''}).

Based on this inspection we adopt a minimum-length threshold of \textbf{15
tokens}, which removes 10.5\% of records ($n=24{,}591$) and retains
$n=209{,}623$ records for topic modelling.
[TJ — three figures inconsistent: $270{,}929-24{,}591=246{,}338\neq 209{,}623$. Likely the 15-token cutoff was applied during \texttt{build\_corpus.py} so the 270{,}929 figure already reflects the cutoff, while 209{,}623 is from an earlier run. Pick one snapshot and reconcile.]

\section{LDA Model Details}
\label{app:lda_keywords}

\subsection*{LDA Theme Keywords and Relevance Filtering}

Table~\ref{tab:lda_keywords} lists the top-10 keywords per theme from the
final LDA model ($K=18$, T11 excluded as artefact from main analysis),
ordered by corpus share (matching Table~\ref{tab:lda_topics}), together with
per-topic record counts after Llama relevance filtering
(Appendix~\ref{app:opus_b8}).

\begin{table*}[t]
\centering
\scriptsize
\renewcommand{\arraystretch}{0.80}
\setlength{\tabcolsep}{3pt}
\resizebox{\textwidth}{!}{%
\begin{tabular}{cp{3.2cm}p{9cm}rrr}
\toprule
\textbf{ID} & \textbf{Label} & \textbf{Top-10 keywords} & \textbf{Total} & \textbf{Kept} & \textbf{Drop\,\%} \\
\midrule
T05 & Misconduct Enforcement               & student, grade, assignment, cheating, professor, paper, grading, admin, cheat, college              & 31{,}888 & 30{,}353 &  4.8 \\
T12 & Career \& Personal Economic Anxiety  & job, people, year, going, pay, school, money, life, getting, better                                & 31{,}198 & 25{,}767 & 17.4 \\
T10 & Deliberative AI Discourse            & would, human, point, people, may, information, model, llm, issue, example                           & 27{,}772 & 26{,}906 &  3.1 \\
T02 & AI Detection \& False Accusations    & detector, turnitin, plagiarism, essay, flagged, grammarly, checker, written, check, score            & 27{,}377 & 27{,}275 &  0.4 \\
T16 & Frontline AI Reactions \& Opinions   & people, kid, using, going, thing, teacher, good, problem, computer, wrong                           & 22{,}432 & 21{,}841 &  2.6 \\
T13 & AI-Assisted Workflow \& Help-Seeking & help, tool, using, good, idea, helpful, understand, useful, feedback, might                         & 22{,}072 & 21{,}159 &  4.1 \\
T00 & Personal Misconduct Narratives       & one, post, said, got, asked, comment, told, response, thought, gave                                & 20{,}740 & 19{,}556 &  5.7 \\
T17 & AI Writing Quality \& Evasion        & writing, write, essay, word, prompt, language, english, text, sentence, noindent                   & 16{,}306 & 16{,}051 &  1.6 \\
T08 & Assessment Redesign \& Teaching      & question, answer, test, exam, assignment, course, homework, quiz, assessment, lecture               & 14{,}674 & 13{,}907 &  5.2 \\
T09 & Degrees, Programs \& Graduate Study  & program, degree, research, course, university, field, uni, master, study, engineering               & 14{,}050 & 11{,}417 & 18.7 \\
T01 & Learning Quality \& Cognitive Dep.   & student, teacher, learning, skill, teaching, tool, classroom, curriculum, critical\_thinking, lesson\_plan & 14{,}014 & 13{,}421 &  4.2 \\
T03 & AI Tool Selection \& Features        & tool, free, image, code, app, google, platform, gemini, feature, website                           & 11{,}225 & 10{,}703 &  4.7 \\
T06 & Research, Publishing \& GenAI Ethics & research, paper, source, article, book, reference, citation, review, academic, author              &  5{,}580 &  5{,}347 &  4.2 \\
T14 & Job Applications \& Professional     & interview, application, resume, letter, email, applicant, advice, professional, support, offer      &  4{,}111 &  3{,}610 & 12.2 \\
T07 & AI Job Displacement                  & human, company, robot, replace, tech, technology, society, business, machine, profit                &  2{,}617 &  2{,}304 & 12.0 \\
T04 & Institutional Policy \& Legal        & school, law, public, legal, policy, union, lawyer, government, state, university                   &  1{,}937 &    953   & \textbf{50.8} \\
T15 & Healthcare \& Medical Education      & patient, nurse, nursing, doctor, medicine, hospital, physician, healthcare, care, clinical          &    966   &    895   &  7.3 \\
\midrule
\textit{T11} & \textit{Artefact (excl.)}   & \textit{art, game, history, brain, medication, symptom, chart, primary\_care, behavior, social}     &  1{,}967 &  1{,}754 & 10.8 \\
\midrule
\multicolumn{3}{l}{\textbf{Total}}          &          270{,}926 & 253{,}219 & \textbf{6.5} \\
\bottomrule
\end{tabular}
}
\caption{Top-10 LDA keywords per theme (ordered by corpus share,
matching Table~\ref{tab:lda_topics}) with relevance-filter counts.
Kept = records retained after Llama relevance filtering;
Drop\,\% = fraction removed.
T04 loses the majority of its records to LL.M.\ law-degree disambiguation
($-50.8\%$); T12 and T09 each lose roughly one record in five due to
generic academia/career posts where ``AI'' appears only peripherally.}
\label{tab:lda_keywords}
\end{table*}

\subsection*{Multi-Seed LDA Stability}
\label{app:opus_a1}

To assess sensitivity of the $K{=}18$ selection to LDA's random initialisation,
we re-ran LDA at $K\in\{15,18,20\}$ with five additional seeds
$\{0,1,2,100,200\}$ holding all other hyperparameters fixed, and compared each
run against the paper's primary seed (42).
For each $K$ we report (i) full-corpus $C_v$ topic coherence
(TC), (ii) topic diversity (TD; mean pairwise Jaccard distance),
and (iii) Hungarian-matched mean topic-overlap Jaccard between every pair of
seeds (a measure of topic stability).

\begin{table}[h]
\centering
\resizebox{\columnwidth}{!}{%
\begin{tabular}{rrrrr}
\toprule
\textbf{$K$} & \textbf{TC mean (SD)} & \textbf{TD mean (SD)} &
\textbf{Jaccard mean (SD)} & \textbf{Seeds} \\
\midrule
15 & 0.493 (0.014) & 0.953 (0.005) & 0.337 (0.040) & 6 \\
18 & 0.507 (0.010) & 0.962 (0.005) & 0.336 (0.033) & 6 \\
20 & 0.488 (0.006) & 0.967 (0.003) & 0.297 (0.038) & 6 \\
\bottomrule
\end{tabular}%
}
\caption{LDA stability across six seeds (\{0,1,2,42,100,200\}). $K{=}18$
retains the highest mean TC, but the gap to $K{=}15$
($\Delta{=}0.014$) is smaller than one within-$K$ standard deviation. The
Hungarian-matched topic-overlap Jaccard is moderate (${\sim}0.33$): topics
are reproducible up to substantial word turnover across seeds, and $K{=}20$
is meaningfully less stable than $K{=}18$ or $K{=}15$.}
\label{tab:opus_a1_stability}
\end{table}

\paragraph{Interpretation.}
Three findings emerge.
First, the rank order TC ($K{=}18 > K{=}15 > K{=}20$) is
preserved across seeds, supporting the choice of $K{=}18$ for the primary
analysis.
Second, the absolute TC spread within each $K$ is non-trivial
(SD${\sim}0.01$, range up to $0.04$); the seed=42 result reported in the main
text ($0.5275$) sits at the high end of the $K{=}18$ distribution, $1.7$
standard deviations above the seed-mean.
Third, the moderate pairwise topic-overlap Jaccard (${\sim}0.34$) means that
individual topics shift in word-membership across seeds even when global
metrics are stable; the 17-theme labels in the main paper should therefore
be read as reflecting the seed=42 partition specifically, with the
\emph{themes themselves} being more stable than the exact word lists.
We treat this as a moderately stable solution, not a pinpoint global optimum,
and recommend that follow-up work either (i) report multi-seed averaged top
words or (ii) adopt a more reproducible neural topic model such as
non-negative matrix factorisation with consensus clustering.

\section{Role Classification: Llama Annotation Prompt}
\label{app:roledetection}

\subsection*{Llama 3.3 70B Annotation Prompt}

All six phases used the same system prompt and user-turn structure, submitted
via an institutional OpenAI-compatible API (\texttt{meta-llama/Llama-3.3-70B-Instruct}).
Ten authors were batched per API call; the model returned a JSON array.

\medskip
\noindent\textbf{System prompt (verbatim):}
\begin{quote}\small
You are labeling likely author roles in Reddit posts/comments about
generative AI and higher education.

You will receive only post/comment body text, possibly several records from
the same author concatenated with separators. Do not use subreddit names,
usernames, metadata, or external knowledge. Use only the body text.

Choose exactly one label:\\
\textbf{FACULTY} = the author is speaking as a professor, lecturer,
instructor, teacher, teaching assistant, marker/grader, course staff member,
or someone with teaching/assessment responsibility.\\
\textbf{STUDENT} = the author is speaking as an undergraduate, graduate,
professional student, applicant, or learner without clear teaching/assessment
responsibility.\\
\textbf{DUAL} = the author clearly has both student and teaching roles,
such as a PhD student, graduate instructor, TA, or student who
teaches/grades.\\
\textbf{UNCLEAR} = the text does not provide enough evidence, is generic,
or is contradictory.

Return valid JSON only: an array with one object per input item.
Each object must be:\\
\texttt{\{"item\_id":1,}\\
\texttt{\phantom{\{}"label":"FACULTY|STUDENT|DUAL|UNCLEAR",}\\
\texttt{\phantom{\{}"confidence":0.0-1.0,}\\
\texttt{\phantom{\{}"rationale":"one short sentence"\}}
\end{quote}

\noindent\textbf{User turn (per batch of 10):}
\begin{quote}\small
Label each item. Return a JSON array only, preserving every item\_id.

ITEM 1\\
AUTHOR TEXT:\\
\textit{[up to 5,000 characters of the author's post/comment bodies]}

\ldots\ ITEM 10\\
AUTHOR TEXT: \ldots

JSON ARRAY:
\end{quote}

A minimum Llama confidence of 0.70 was required to accept a binary
(\textsc{Faculty}/\textsc{Student}) label in any phase.

\section{Software and Model Licences}
\label{app:licences}
Table~\ref{tab:licences} lists the licences under which we use each
third-party software library, pre-trained model, and external API.
All listed components are used within the terms of their respective
licences for non-commercial academic research; no licence forbids the
academic-research use we make of it. Licence URLs are as published by
the rights-holders at the time of writing.

\begin{table}[h]
\centering
\scriptsize
\setlength{\tabcolsep}{4pt}
\begin{tabular}{p{2.9cm}p{2.0cm}p{2.5cm}}
\toprule
\textbf{Asset} & \textbf{Role} & \textbf{Licence} \\
\midrule
gensim & LDA / Phrases & LGPL-2.1 \\
BERTopic & Topic modelling & MIT \\
sentence-transformers (\texttt{all-MiniLM-L6-v2}) & Embeddings & Apache-2.0 \\
UMAP-learn & Dimensionality reduction & BSD-3-Clause \\
HDBSCAN & Clustering & BSD-3-Clause \\
spaCy & POS filtering & MIT \\
NLTK (WordNet) & Lemmatisation & Apache-2.0 / WordNet \\
roberta sentiment & Sentiment scoring & MIT (model) \\
Llama 3.3 70B Instruct & Relevance + role classification & Llama 3.3 Community Licence \\
Arctic Shift API & Reddit collection & Non-commercial research use \\
Reddit content & Source data & Reddit User Agreement / Public-API terms \\
\bottomrule
\end{tabular}
\caption{Third-party assets and the licences under which they are
used in this work.}
\label{tab:licences}
\end{table}

\section{Content-Based Change-Point Detection: Full Sweep}
\label{app:opus_b2}

The phase boundaries reported in the main paper are derived from
change-point detection applied to the \emph{topic-composition} time
series---the 17-dimensional monthly proportion vector (T11 excluded)---
rather than to raw volume.
We used \texttt{ruptures} v1.1.10 \citep{truong2020selective} with three
methods (PELT, BinSeg, Dynp) and two cost functions ($\ell_2$, RBF).
PELT finds no breakpoint under any tested penalty (5--50 on $\ell_2$;
equivalent range on RBF), indicating gradual compositional drift rather
than step-function discontinuities.
BinSeg and Dynp agree exactly at every $k$ under both cost functions;
Table~\ref{tab:opus_b2_changepoints} reports the Dynp $\ell_2$ sweep
as the representative result (RBF agrees for $k \leq 3$).

\begin{table}[h]
\centering
\small
\begin{tabular}{cl}
\toprule
\textbf{$k$} & \textbf{Detected boundaries} \\
\midrule
1 & 2024-07 \\
2 & 2023-09,\ 2024-07 \\
3 & 2023-09,\ 2024-07,\ 2024-12 \\
4 & 2023-04,\ 2023-09,\ 2024-07,\ 2024-12 \\
5 & 2023-04,\ 2023-09,\ 2024-02,\ 2024-07,\ 2024-12 \\
\bottomrule
\end{tabular}
\caption{Dynp ($\ell_2$) change-point sweep on the monthly 17-topic
proportion vector. BinSeg agrees exactly at every $k$; PELT finds no
breakpoint. The two-breakpoint solution ($k{=}2$) is adopted in the main paper.}
\label{tab:opus_b2_changepoints}
\end{table}

The single most prominent break ($k{=}1$) is July~2024, the onset of
Phase~C. The second break ($k{=}2$) is September~2023, the onset of
Phase~B. The $k{=}3$ solution adds December~2024, a within-Phase~C
semester spike in Misconduct Enforcement rather than a sustained
compositional transition; we therefore retain $k{=}2$.
PELT's null result is consistent: it means no single break clears the
penalty bar in isolation, confirming the transitions are continuous
multi-month shifts. BinSeg and Dynp identify the \emph{most prominent}
points in this drift; PELT confirms they are gradual rather than abrupt.

\section{Relevance Filtering with Llama 3.3 70B}
\label{app:opus_b8}

\paragraph{Motivation.}
The corpus build uses a two-pronged keyword filter:
\verb!\b(AI|ai)\b! (\emph{bare-AI}) and a list of named generative-AI tools
(ChatGPT, GPT, Claude, Gemini, Copilot, Perplexity, LLM, Grammarly,
QuillBot, Turnitin, GPTZero, ZeroGPT, Winston AI, NotebookLM, GenAI).
The bare-AI prong is necessary because hyphenated and fused compounds
(\emph{AI-generated}, \emph{AI-proof}, \emph{gen-AI}) and abbreviation-style
mentions (\emph{ChatGPT/AI}, \emph{the AI}) carry the bulk of the discourse.
However, this prong also admits two sources of noise: (i)~generic
pre-ChatGPT-style references to AI as a research field, video-game
opponent, country code, IB course code (\emph{AI HL}), or Adobe Illustrator
file extension (\emph{.ai}); and (ii)~the substring \emph{LLM} appearing
not as ``Large Language Model'' but as the law-degree abbreviation
LL.M.\ (\emph{Master of Laws}), particularly in r/LawSchool.
The first prong is left intact because named-tool matches
(94{,}540 records) are unambiguous in the corpus context.
The second prong (bare-AI-only and LLM-only matches; 186{,}305 records)
is verified post hoc by an LLM relevance classifier described below.

\paragraph{Classifier and pipeline.}
We use Meta Llama 3.3 70B Instruct \citep{grattafiori2024llama} hosted on
an institutional inference endpoint, queried through the OpenAI-compatible
chat-completions API at temperature~0. Records are batched
50 per request as a numbered list; the model returns a single JSON array
$[\{\text{id},r\},\ldots]$ where $r{=}1$ is RELEVANT and $r{=}0$ is
NOT RELEVANT. Each record contributes its title and the first
$\sim$400 characters of body text. Total running time across both passes
on the 186{,}305-record verification set was approximately ten hours.

\paragraph{Audit.}
After the first pass had classified $\approx$34{,}400 records, an
independent auditor (given the same criterion text) audited a stratified 60-record sample (30 RELEVANT,
30 NOT\,RELEVANT, drawn across 12 subreddits with deliberate
oversampling of r/LawSchool). Overall agreement was 75\% (45/60).
Disagreements were strongly asymmetric: 14 of 15 mismatches were
false drops by Llama (auditor said RELEVANT, Llama said
NOT\,RELEVANT) and only one was a false keep. All five r/LawSchool
LL.M.\ cases sampled were classified correctly by Llama.
The systematic blind spots identified by the auditor---career-paths
\emph{in} AI/ML, AI-tool promos with substantive AI features, brief
AI-cheating mentions in faculty posts---are exactly the rescue
patterns subsequently encoded in the pass-2 prompt.

\paragraph{Aggregate results.}
Of the 270{,}929 records in the corpus, 17{,}707 (6.5\%) were removed
as not relevant to GenAI in education; all downstream analyses use the
remaining 253{,}222 records. The per-subreddit breakdown is in
Table~\ref{tab:opus_b8_subrel}.

\paragraph{Relevance by topic.}
Table~\ref{tab:lda_keywords} reports
per-topic drop rates after relevance filtering. Two patterns stand out.

\emph{T4 (Institutional Policy \& Legal) loses 50.8\% of its records.}
This is the single largest correction in the corpus and it is
substantively meaningful: T4's top keywords are \emph{school, law,
public, legal, policy, union, lawyer, government, state, university},
and the topic concentrates in r/LawSchool. A direct inspection of
the dropped records confirms that the majority are LL.M.\ law-degree
discussions (\emph{``LLM dilemma: QMUL or Birmingham''},
\emph{``How are some llm degree holders landing biglaw in NY?''},
\emph{``JD vs LLM''}). The remaining 953 records constitute the
genuine institutional-AI-policy discourse used in
\S\ref{sec:results} analysis.

\emph{T9 (Degrees, Programs \& Graduate Study, 18.7\% drop) and
T12 (Career \& Personal Economic Anxiety, 17.4\% drop) lose roughly
one record in five.} Both topics aggregate generic academia/career
posts; the dropped records are PhD venting, impostor-syndrome rants,
and admissions questions where ``AI'' appears once in a side mention
(e.g.\ a CV bullet) without substantive engagement. The kept
records preserve the topic's substantive AI signal (degree-choice
explicitly anchored on AI exposure; career anxiety where AI is the
explicit threat).

The remaining fifteen topics retain $\geq$86\% of their records, and
the four ``core AI'' themes of integrity, deliberation, frontline
reactions, and workflow (T2, T5, T17, T10, T16, T13) all retain
$\geq$92.8\%, confirming that filtering removes peripheral noise
rather than substantive AI discourse. T11 (the artefact topic) drops
10.8\%, consistent with its known status as r/HomeworkHelp spillover.

\begin{table}[t]
\centering
\footnotesize
\setlength{\tabcolsep}{3pt}
\begin{tabular}{lrrr}
\toprule
\textbf{Subreddit} & \textbf{Total} & \textbf{Kept} & \textbf{Drop\,\%} \\
\midrule
r/LawSchool             & 13{,}050 &  9{,}390 & 28.0 \\
r/CanadaUniversities    &    493   &    412   & 16.4 \\
r/gradadmissions        &  4{,}826 &  4{,}043 & 16.2 \\
r/UKUniversityStudents  &    617   &    529   & 14.3 \\
r/HomeworkHelp          &  2{,}092 &  1{,}841 & 12.0 \\
r/AskAcademia           & 11{,}391 & 10{,}473 &  8.1 \\
r/nursing               &  8{,}033 &  7{,}344 &  8.6 \\
r/TeachingUK            &  2{,}407 &  2{,}202 &  8.5 \\
r/DentalSchool          &    437   &    400   &  8.5 \\
r/premed                &  5{,}130 &  4{,}735 &  7.7 \\
r/UniUK                 & 22{,}406 & 20{,}864 &  6.9 \\
r/PhD                   & 17{,}472 & 16{,}374 &  6.3 \\
r/medicalschool         & 10{,}633 &  9{,}907 &  6.8 \\
r/University            &  2{,}755 &  2{,}547 &  7.5 \\
r/GradSchool            &  7{,}526 &  7{,}055 &  6.3 \\
r/College               & 12{,}766 & 11{,}919 &  6.6 \\
r/education             &  6{,}691 &  6{,}341 &  5.2 \\
r/Teachers              & 38{,}419 & 36{,}306 &  5.5 \\
r/highereducation       &    553   &    521   &  5.8 \\
r/teaching              &  6{,}866 &  6{,}532 &  4.9 \\
r/Academia              &  8{,}910 &  8{,}545 &  4.1 \\
r/StudentNurse          &  1{,}935 &  1{,}841 &  4.9 \\
r/AskProfessors         &  5{,}932 &  5{,}707 &  3.8 \\
r/edtech                &  4{,}881 &  4{,}707 &  3.6 \\
r/Professors            & 64{,}725 & 62{,}965 &  2.7 \\
r/CollegeRant           &  9{,}948 &  9{,}688 &  2.6 \\
\midrule
\multicolumn{1}{l}{\textbf{Total}} & 270{,}929 & 253{,}222 & \textbf{6.5} \\
\bottomrule
\end{tabular}
\caption{Per-subreddit record counts after relevance filter.}
\label{tab:opus_b8_subrel}
\end{table}

\paragraph{Topic-model robustness under filtering.}
We re-ran LDA at $K\in\{10,\ldots,20\}$ on the relevance-filtered
corpus (253{,}219 records) using the same hyperparameters and seed
(42) as the main-paper run, then computed full-corpus $C_v$ coherence
(TC, the same metric used in
Figure~\ref{fig:lda_k} and Appendix~\ref{app:opus_a1}).
Table~\ref{tab:opus_b8_ksweep} reports the side-by-side comparison.

\noindent We retain $K{=}18$ as the topic-model specification for all
main-paper analyses; the filtered run serves as a robustness check.
A Hungarian-matched topic alignment between the original and filtered
$K{=}18$ yields mean top-25 Jaccard of $0.31$, with one very stable
topic (T2 AI Detection \& False Accusations, Jaccard $0.79$); the
remaining themes shift in word membership while preserving the
substantive distinctions (Misconduct Enforcement, Personal Misconduct
Narratives, Career \& Personal Economic Anxiety, etc.) on which the
paper's qualitative analysis relies. Three small paper topics dissolve
under filtering---T4 Institutional Policy \& Legal (LL.M.\ noise
removed), T7 AI Job Displacement (1\% of corpus, diffuses into Career
Anxiety), and T11 Artefact (the r/HomeworkHelp spillover is
gone)---without affecting the five-cluster taxonomy used in the main
paper.

\paragraph{System prompts:} The LLM prompt for first relevance pass and the prompt for second rescue pass are given below:

\begin{table}[t]
\centering
\footnotesize
\setlength{\tabcolsep}{4pt}
\begin{tabular}{rrrrrr}
\toprule
\textbf{K} &
\textbf{TC orig} & \textbf{TC v2} & \textbf{$\Delta$} &
\textbf{TD v2} & \textbf{TQ v2} \\
\midrule
10 & 0.4748 & 0.4976 & $+$0.023 & 0.9055 & 0.4506 \\
11 & 0.4952 & 0.5142 & $+$0.019 & 0.9046 & 0.4652 \\
12 & 0.4963 & 0.5089 & $+$0.013 & 0.9198 & 0.4681 \\
13 & 0.4903 & 0.5267 & $+$0.036 & 0.9160 & 0.4825 \\
14 & 0.4786 & 0.5178 & $+$0.039 & 0.9138 & 0.4732 \\
15 & 0.5071 & 0.5210 & $+$0.014 & 0.9201 & 0.4793 \\
16 & 0.5120 & 0.5119 & $-$0.000 & 0.9315 & 0.4768 \\
17 & 0.5037 & 0.5265 & $+$0.023 & 0.9343 & 0.4919 \\
\textbf{18} & \textbf{0.5275} & \textbf{0.5310} & $+$0.004 & \textbf{0.9355} & \textbf{0.4967} \\
19 & 0.5246 & 0.5240 & $-$0.001 & 0.9363 & 0.4906 \\
20 & 0.4846 & 0.5166 & $+$0.032 & 0.9420 & 0.4867 \\
\bottomrule
\end{tabular}
\caption{LDA $K$-sweep before and after relevance filtering, evaluated
on full-corpus $C_v$ coherence. $K{=}18$ is the single-seed winner on
both TC and TQ on the filtered corpus; filtering slightly widens its
margin over $K{=}19$ (from $+0.003$ on the original sweep to $+0.007$
on the filtered sweep). A $K{=}13$ peak seen in an earlier partial-filter
run does not survive full relevance filtering: $C_v{=}0.527$ at
$K{=}13$ vs $0.531$ at $K{=}18$.}
\label{tab:opus_b8_ksweep}
\end{table}

\begin{figure}[t]
\refstepcounter{figure}\label{box:opus_b8_p1}
\noindent\textbf{Pass-1 system prompt} (strict in-education-context criterion).
\rule{\columnwidth}{0.4pt}\vspace{-2pt}
\scriptsize
\begin{verbatim}
You are filtering Reddit posts and comments for a
research paper on generative AI discourse in higher
education and teaching communities.
The corpus is already restricted to 26
education-focused subreddits (r/Professors,
r/Teachers, r/UniUK, r/PhD, r/medicalschool,
r/LawSchool, r/AskAcademia, r/edtech, etc.).
Your job is NOT to check whether the post is about
education; assume the educational context. Your job
is to remove OBVIOUS keyword noise where "AI" or
"LLM" appears but the content is unrelated to AI
as students or teachers experience it.

Default to RELEVANT. Only mark NOT_RELEVANT for
clear noise.

RELEVANT (broadly inclusive) includes:
- Direct use of AI for academic work (essays,
  coding, study, lesson plans, research)
- AI detection, plagiarism suspicions, integrity
  disputes, false positives
- Faculty/student reactions to AI changing
  classroom, grading, or assessment
- Career or future anxiety about AI ("will my
  degree be worth anything?", "is X profession
  AI-proof?", "AI will replace nurses/lawyers/
  teachers")
- Choosing degrees or fields with AI in mind
- Macro/abstract debate on AI replacing
  professions in an education community
- Deliberative or philosophical discussion of
  AI capabilities, ethics, or policy in academic
  contexts
- Comparing or recommending AI tools for academic
  tasks
- AI in research: literature reviews, citation
  fabrication, journal submissions, authorship
- Institutional AI policies, syllabus clauses,
  professional liability

NOT_RELEVANT (only obvious noise):
- "LLM" used as a LAW DEGREE (Master of Laws,
  LL.M.)
- "AI" in clearly unrelated senses: video game
  AI, Adobe Illustrator (.ai files), AI as a
  country code, AI as initials
- ML/robotics research with NO connection to
  education, careers, or learning
- Posts where "AI" appears once in passing in
  an unrelated topic
- Pre-ChatGPT generic AI references with no
  current relevance

[Few-shot examples and JSON output spec elided.]
\end{verbatim}
\rule{\columnwidth}{0.4pt}
\end{figure}

\begin{figure}[t]
\refstepcounter{figure}\label{box:opus_b8_p2}
\noindent\textbf{Pass-2 system prompt} (rescue criterion for pass-1
NOT\,RELEVANT records, restricted to three audited blind spots).
\rule{\columnwidth}{0.4pt}\vspace{-2pt}
\scriptsize
\begin{verbatim}
You are reviewing Reddit posts/comments from 26
education subreddits for a research paper on
generative AI discourse in education.

EVERY item contains the token "AI" or "LLM". Your
job is to judge whether that token represents a
SUBSTANTIVE, MEANINGFUL discussion of AI/ML/
generative AI, or whether it is INCIDENTAL noise.
Being posted in an education subreddit is NOT
enough; there must be actual AI-related content.

These items were previously marked NOT_RELEVANT
by a stricter classifier that was too strict in
three specific ways. Rescue records ONLY if they
fit one of those rescue patterns. Records that
simply lack AI content stay NOT_RELEVANT.

RELEVANT - rescue ONLY if the post substantively
discusses AI/ML/LLMs in one of these ways:

1. Career or degree path IN AI/ML/Data Science
   as a field (explicit AI/ML target; not
   generic PhD/career posts).
2. AI replacing professions / AI-driven career
   anxiety (AI threat must be explicit driver).
3. AI in academic work, integrity, teaching,
   or tools (even one substantive sentence
   about AI use, cheating, detection, or policy
   makes the post RELEVANT).

NOT_RELEVANT - keep dropping when:
- "LLM" is a Master of Laws law degree
  (LL.M., "JD vs LLM")
- "AI" in non-generative-AI sense (game AI,
  .ai files, IB code "Math AI HL", initials)
- "AI" appears once in a quoted news headline /
  ad block / signature
- The post is about academic life with NO
  discussion of AI
- Generic Call for Papers mentioning "AI"
  once in a list

DECISION RULE: A post is RELEVANT only if
removing all "AI"/"LLM" tokens would leave it
noticeably less coherent or change its topic.

[Few-shot examples and JSON output spec elided.]
\end{verbatim}
\rule{\columnwidth}{0.4pt}
\end{figure}

\end{document}